\documentclass{amsart}
\usepackage{amsmath}
\usepackage{amssymb}

\theoremstyle{plain}
\newtheorem{theorem}{Theorem}[section]
\newtheorem{lemma}[theorem]{Lemma}

\theoremstyle{definition}
\newtheorem{definition}[theorem]{Definition}

\theoremstyle{remark}
\newtheorem{remark}[theorem]{Remark}
\numberwithin{equation}{section}

\input{tcilatex}

\begin{document}
\title[De Rham-Hodge-Skrypnik theory ]{A survey of the spectral and
differential geometric aspects of the De Rham-Hodge-Skrypnik theory related
with Delsarte transmutation operators in multidimension and its applications
to spectral and soliton problems. Part 1}
\thanks{The third author was supported in part by a local AGH grant.}
\author{Y.A. Prykarpatsky*)**)}
\address{*)The AMM University of Science and Technology, Department of
Applied Mathematics, Krakow 30059 Poland, and Brookhaven Nat. Lab., CDIC,
Upton, NY, 11973 USA}
\email{yarchyk@imath.kiev.ua, yarpry@bnl.gov}
\author{A.M. Samoilenko**)}
\address{**)The Institute of Mathematics, NAS, Kyiv 01601, Ukraine}
\author{A. K. Prykarpatsky***)}
\address{***)The AMM University of Science and Technology, Department of
Applied Mathematics, Krakow 30059 Poland}
\email{pryk.anat@ua.fm, prykanat@cybergal.com}
\date{2004}

\begin{abstract}
A review on spectral and differential-geometric properties of Delsarte
transmutation operators in multidimension is given. Their differential
geometrical and topological structure in multidimension is analyzed, the
relationships with De Rham-Hodge-Skrypnik theory of generalized differential
complexes are stated. Some applications to integrable dynamical systems
theory in multidimension are presented.
\end{abstract}

\maketitle


\section{Spectral operators and generalized eigenfunctions expansions}

\setcounter{equation}{0}1.1. Let $\mathcal{H}$ be a Hilbert space in which
there is defined a linear closable operator $L\in \mathcal{L}(\mathcal{H})$
with a dense domain $D(L)\subset \mathcal{H}.$ Consider the standard
quasi-nucleous Gelfand rigging \cite{Be} of this Hilbert space $\mathcal{H}$
with corresponding positive $\mathcal{H}_{+}$ and negative $\mathcal{H}_{-}$
Hilbert spaces as follows:

\begin{equation}
D(L)\subset \mathcal{H}_{+}\subset \mathcal{H}\subset \mathcal{H}_{-}\subset
D^{\prime }(L),  \label{0}
\end{equation}%
being suitable for proper analyzing the spectral properties of the operator $%
L$ in $\mathcal{H}$. We shall use below the following definition motivated
by considerations from \cite{Be}, Section 5.

\begin{definition}
An operator $L\in \mathcal{L}(\mathcal{H})$ will be called \textbf{spectral}
if for all Borel subsets $\Delta \subset \sigma (L)$ of the spectrum $\sigma
(L)\subset \mathbb{C}$ and for all pairs $(u,v)\in \mathcal{H}_{+}\times 
\mathcal{H}_{+}$ there\ are defined the following expressions:
\end{definition}

\begin{equation}
L=\int_{\sigma (L)}\lambda \mathrm{d}\mathrm{E}(\lambda ),\ \ (u,\mathrm{E}%
(\Delta )v)=\int_{\Delta }(u,\mathrm{P}(\lambda )v)\mathrm{d}\rho _{\sigma
}(\lambda ),  \label{eq:1}
\end{equation}%
where $\rho _{\sigma }$ is some finite Borel measure on the spectrum $\sigma
(L)$, $\mathrm{E}$ is some self-adjoint projection operator measure on the
spectrum $\sigma (L)$, such that $\ \mathrm{E}(\Delta )\mathrm{E}(\Delta
^{\prime })=\mathrm{E}(\Delta \cap \Delta ^{\prime })$ for any Borel subsets 
$\Delta ,\Delta ^{\prime }\subset \sigma (L),$ and $\ \mathrm{P}(\lambda ):\ 
\mathcal{H}_{+}\rightarrow \mathcal{H}_{-},$ $\lambda \in \sigma (L),$ is
the corresponding family of nucleous integral operators from $\mathcal{H}%
_{+} $ into $\mathcal{H}_{-}.$

As a consequence of the expression (\ref{eq:1}) one can write down that
formally in the weak topology of $\mathcal{H}$

\begin{equation}
\mathrm{E}(\Delta )=\int_{\Delta }\mathrm{P}(\lambda )\mathrm{d}\rho
_{\sigma }(\lambda )  \label{eq:2}
\end{equation}%
for any Borel subset $\Delta \subset \sigma (L).$

Similarly to (\ref{eq:1}) and (\ref{eq:2}) can write down the corresponding
expressions for the adjoint spectral operator $L^{\ast }\in \mathcal{L}(%
\mathcal{H})$ whose domain $\mathrm{D}(L^{\ast })\subset \mathcal{H}$ is
assumed to be also dense in $\mathcal{H}:$ 
\begin{equation}
(\mathrm{E}^{\ast }(\Delta )u,v)=\int_{\Delta }(P^{\ast }(\lambda )u,v)%
\mathrm{d}\rho _{\sigma }^{\ast }(\lambda ),  \label{eq:3}
\end{equation}%
\begin{equation*}
\mathrm{E}^{\ast }(\Delta )=\int_{\Delta }P^{\ast }(\lambda )\mathrm{d}\rho
_{\sigma }^{\ast }(\lambda ),
\end{equation*}%
where $\mathrm{E}^{\ast }$ is the corresponding projection spectral measure
on Borel subsets $\Delta \in \sigma (L^{\ast }),$ $\mathrm{P}^{\ast
}(\lambda ):$ $\mathcal{H}\ \rightarrow \ \mathcal{H},$ $\lambda \in \sigma
(L^{\ast }),$ is the corresponding family of nucleous integral operators in $%
\mathcal{H}$ and $\rho _{\sigma }^{\ast }$ is some finite Borel measure on
the spectrum $\sigma (L^{\ast }).$ We will assume, moreover, that the
following conditions 
\begin{equation}
\mathrm{P}(\mu )(L-\mu \mathrm{I})v=0,\ \ \ \mathrm{P}^{\ast }(\lambda
)(L^{\ast }-\bar{\lambda}\mathrm{I})u=0  \label{eq:4}
\end{equation}
hold for all $u\in \mathrm{D}(L^{\ast }),$ $v\in \mathrm{D}(L),$ where $\bar{%
\lambda}\in \sigma (L^{\ast }),\ \mu \in \sigma (L).$ In particular, one
assumes also that $\sigma (L^{\ast })=\bar{\sigma}(L)$.

1.2. Proceed now to a description of the corresponding to operators $L$ and $%
L^{\ast }$ generalized eigenfunctions via the approach devised in \cite{Be}.
\ We shall speak that an operator $L\in \mathcal{L}(\mathcal{H})$ with a
dense domain $\mathrm{D}(L)$ allows a rigging continuation, if one can find
another dense in $\mathcal{H}_{+}$ topological subspace $\mathrm{D}%
_{+}(L^{\ast })\subset \mathrm{D}(L^{\ast }),$ such that the adjoint
operator $L^{\ast }\in \mathcal{L}(\mathcal{H})$ maps it continuously into $%
\mathcal{H}_{+}.$

\begin{definition}
A vector $\psi _{\lambda }\in \mathcal{H}_{-}$ is called a generalized
eigenfunction of the operator $L\in \mathcal{L}(\mathcal{H})$ corresponding
to an eigenvalue $\lambda \in \sigma (L)$ if 
\begin{equation}
((L^{\ast }-\bar{\lambda}\mathrm{I})u,\psi _{\lambda })=0  \label{eq:5}
\end{equation}%
for all $u\in \mathrm{D}_{+}(L^{\ast }).$
\end{definition}

It is evident that in the case when $\psi _{\lambda }\in \mathrm{D}(L),\ \
\lambda \in \sigma (L),$ then $L\psi _{\lambda }=\lambda \psi _{\lambda }$
as usually. The definition (\ref{eq:5}) is related \cite{Be} with some
extension of the operator $L:\mathcal{H}\ \mapsto \ \mathcal{H}$. Since the
operator $L^{\ast }:\ \mathrm{D}_{+}(L^{\ast })\ \rightarrow \ \mathcal{H}%
_{+}$ is continuous one can define the adjoint operator $L_{ext}:=\ L^{\ast
,+}:\ \mathcal{H}_{-}\ \rightarrow \ \mathrm{D}_{-}(L^{\ast })$ with respect
to the standard scalar product in $\mathcal{H},$ that is 
\begin{equation}
(L^{\ast }v,u)=(v,L^{\ast ,+}u)  \label{eq:6}
\end{equation}%
for any $v\in \mathrm{D}_{+}(L^{\ast })$ and $u\in \mathcal{H}_{-}$ and
coinciding with the operator $L:\mathcal{H}\ \rightarrow \ \mathcal{H}$ upon 
$\mathrm{D}(L).$ Now the definition (\ref{eq:5}) of a generalized
eigenfunction $\psi _{\lambda }\in \mathcal{H}_{-}$ for $\lambda \in \sigma
(L)$ is equivalent to the standard expression 
\begin{equation}
L_{ext}\psi _{\lambda }=\lambda \psi _{\lambda }.  \label{eq:7}
\end{equation}%
If to define the scalar product 
\begin{equation}
(u,v):=\ (u,v)_{+}+(L^{\ast }u,L^{\ast }v)_{+}  \label{eq:8}
\end{equation}%
on the dense subspace $\mathrm{D}_{+}(L^{\ast })\subset \mathcal{H}_{+},$
then this subspace can be transformed naturally into the Hilbert space $%
\mathrm{D}_{+}(L^{\ast }),$ whose adjoint \textquotedblright
negative\textquotedblright\ space $\mathrm{D}_{+}^{\prime }(L^{\ast }):=%
\mathrm{D}_{-}(L^{\ast })\supset \mathcal{H}_{-}.$ Take now any generalized
eigenfunction $\psi _{\lambda }\in Im\ \mathrm{P}(\lambda )\subset \mathcal{H%
}_{-},\ \lambda \in \sigma (L),$ of the operator $L:\mathcal{H}\ \rightarrow
\ \mathcal{H}.$ Then, as one can see from (\ref{eq:4}), $L_{ext}^{\ast
}\varphi _{\lambda }=\bar{\lambda}\varphi _{\lambda }$ for some function $%
\varphi _{\lambda }\in Im\ \mathrm{P}^{\ast }(\lambda )\subset \mathcal{H}%
_{-},$ $\bar{\lambda}\in \sigma (L^{\ast }),$ and $L_{ext}^{\ast }:\mathcal{H%
}_{-}\rightarrow $ $\mathrm{D}_{-}(L)$ is the corresponding extension of the
adjoint operator $L^{\ast }:\mathcal{H}\ \rightarrow \ \mathcal{H}$ by means
of reducing, as above, the domain $\mathrm{D}(L)$ to a new dense in $%
\mathcal{H}_{+}$ domain $\mathrm{D}_{+}(L)\subset \mathrm{D}(L)$ on which
the operator $L:\mathrm{D}_{+}(L)\rightarrow \mathcal{H}_{+\text{ }}$is
continuous$.$

\section{Semi-linear forms, generalized kernels and congruence of operators}

2.1. Let us consider any continuous semi-linear form $\mathrm{K}:\mathcal{H}%
\times \mathcal{H}\ \rightarrow \ \mathbb{C}$ in a Hilbert space $\mathcal{H}%
.$ The following classical theorem holds.

\begin{theorem}
(\textbf{L. Schwartz}; see \cite{Be} ) Consider a standard Gelfand rigged
chain of Hilbert spaces (\ref{0}) which is, as usually, invariant under the
complex involution $\mathbb{C}:\rightarrow \mathbb{C}^{\ast }.$ Then any
continuous semi-linear form $\mathrm{K}:\mathcal{H}\times \mathcal{H}\
\rightarrow \ \mathbb{C}$ \ can be written down by means of a generalized
kernel $\mathrm{\hat{K}}\in \mathcal{H}_{-}\otimes \mathcal{H}_{-}$ as
follows: 
\begin{equation}
\mathrm{K}[u,v]=(\mathrm{\hat{K}},v\otimes u)_{\mathcal{H}\otimes \mathcal{H}%
}  \label{eq:9}
\end{equation}%
for any $u,v\in \mathcal{H}_{+}\subset \mathcal{H}.$ The kernel $\mathrm{%
\hat{K}}\in \mathcal{H}_{-}\otimes \mathcal{H}_{-}$ allows the
representation 
\begin{equation*}
\mathrm{\hat{K}}=(\mathrm{D}\otimes \mathrm{D})\bar{K},
\end{equation*}%
where $\mathrm{\bar{K}}\in \mathcal{H}\otimes \mathcal{H}$ \ is a usual
kernel and $\mathrm{D}:\mathcal{H}\ \rightarrow \ \mathcal{H}_{-}$ is the
square root $\sqrt{\mathrm{J}^{\ast }}$ from a positive operator $\mathrm{J}%
^{\ast }:\mathcal{H}\ \rightarrow \ \mathcal{H}_{-},$ being a
Hilbert-Schmidt embedding of $\mathcal{H}_{+}$ into $\mathcal{H}$ \ with
respect to the chain (\ref{0}). Moreover, the related kernels $(\mathrm{D}%
\otimes \mathrm{I})\mathrm{\bar{K}}$, $(\mathrm{I}\otimes \mathrm{D})\mathrm{%
\bar{K}}\in \mathcal{H}\times \mathcal{H}$ are usual ones too.
\end{theorem}

Take now, as before, an operator $L:\mathcal{H}\rightarrow \mathcal{H}$ with
a dense domain $\mathrm{D}(L)\subset \mathcal{H}$ allowing the Gelfand
rigging continuation (\ref{0}) introduced in the preceding Section. Denote
also by $\mathrm{D}_{+}(L^{\ast })\subset \mathrm{D}(L^{\ast })$ the related
dense in $\mathcal{H}_{+}$ subspace.

\begin{definition}
A set of generalized kernels $\mathrm{\hat{Z}}_{\lambda }\subset \mathcal{H}%
_{-}\otimes \mathcal{H}_{-}$ \ for $\lambda \in \sigma (L)\cap \bar{\sigma}%
(L^{\ast })$ will be called \textbf{elementary } concerning the operator $L:%
\mathcal{H}\rightarrow \mathcal{H}$ \ if for any $\lambda \in \sigma (L)\cap 
\bar{\sigma}(L^{\ast }),$ the norm $||\mathrm{\hat{Z}}_{\lambda }||_{%
\mathcal{H}_{-}\otimes \mathcal{H}_{-}}<\infty $ and
\end{definition}

\begin{equation}
(\mathrm{\hat{Z}}_{\lambda },((\Delta -\lambda \mathrm{I})v)\otimes u)=0,%
\text{ \ }(\mathrm{\hat{Z}}_{\lambda },v\otimes (\mathbb{L}^{\ast }-\lambda
\ \mathrm{I})u)=0  \label{eq:10}
\end{equation}%
for all $(u,v)\in \mathcal{H}_{-}\otimes \mathcal{H}_{-}.$

2.2. Assume further, as above, that all our functional spaces are invariant
with respect to the involution $\mathbb{C}:\rightarrow \mathbb{C}^{\ast }$
and put $\mathrm{D}_{+}:=\mathrm{D}_{+}(L^{\ast })=\mathrm{D}_{+}(L)\subset 
\mathcal{H}_{+}.$ Then one can build the corresponding extensions $%
L_{ext}\supset L$ and $L_{ext}^{\ast }\supset L^{\ast },$ being linear
operators continuously acting from $\mathcal{H}_{-}$ into $\mathcal{D}_{-}:=%
\mathrm{D}_{+}^{\prime }.$ The chain (\ref{0}) is now extended to the chain 
\begin{equation}
\mathrm{D}_{+}\subset \mathcal{H}_{+}\subset \mathcal{H}\subset \mathcal{H}%
_{-}\subset \mathrm{D}_{-}  \label{eq:11a}
\end{equation}%
and is assumed also that the unity operator $\mathrm{I}:\mathcal{H}_{-}\
\rightarrow \ \mathcal{H}_{-}\subset \mathrm{D}_{-}$ is extended naturally
as the imbedding operator from $\mathcal{H}_{-}$ into $\mathrm{D}_{-}.$ Then
equalities (\ref{eq:10}) can be equivalently written down \cite{Be} as
follows: 
\begin{equation}
(L_{ext}\otimes \mathrm{I})\ \mathrm{\hat{Z}}_{\lambda }=\lambda \mathrm{%
\hat{Z}}_{\lambda },\text{ \ }(\mathrm{I}\otimes L_{ext}^{\ast })\ \mathrm{%
\hat{Z}}_{\lambda }=\lambda \mathrm{\hat{Z}}_{\lambda }  \label{eq:11}
\end{equation}%
for any $\lambda \in \sigma (L)\cap \bar{\sigma}(L^{\ast }).$ Take now a
kernel $\mathrm{\hat{K}}\in \mathcal{H}_{-}\otimes \mathcal{H}_{-}$ and
suppose that the following operator equality 
\begin{equation}
(L_{ext}\otimes \mathrm{I})\ \mathrm{\hat{K}}=(\mathrm{I}\otimes
L_{ext}^{\ast })\ \mathrm{\hat{K}}  \label{eq:12}
\end{equation}%
holds. Since the equation (\ref{eq:11}) can be written down in the form 
\begin{equation}
(L_{ext}\otimes \mathrm{I})\ \mathrm{\hat{Z}}_{\lambda }=(\mathrm{I}\otimes
L_{ext}^{\ast })\ \mathrm{\hat{Z}}_{\lambda }  \label{eq:13}
\end{equation}%
for any $\lambda \in \sigma (L)\cap \bar{\sigma}(L^{\ast }),$ the following
characteristic theorem \cite{Be} holds.

\begin{theorem}
(see \cite{Be}, chapter 8, p.621) \ Let a kernel $\mathrm{\hat{K}}\in 
\mathcal{H}_{-}\otimes \mathcal{H}_{-}$ satisfy the condition (\ref{eq:12}).
Then due to (\ref{eq:13}) there exists such a finite Borel measure $\rho
_{\sigma }$ defined on Borel subsets $\Delta \subset \sigma (L)\cap \bar{%
\sigma}(L^{\ast }),$ that the following weak spectral representation 
\begin{equation}
\mathrm{\hat{K}}=\int_{\sigma (L)\cap \bar{\sigma}(L^{\ast })}\mathrm{\hat{Z}%
}_{\lambda }\mathrm{d}\rho _{\sigma }(\lambda )  \label{eq:14}
\end{equation}%
holds. Moreover, due to (\ref{eq:11}) one can write down the following
representation 
\begin{equation*}
\mathrm{\hat{Z}}_{\lambda }=\psi _{\lambda }\otimes \varphi _{\lambda },
\end{equation*}%
where $L_{ext}\ \psi _{\lambda }=\ \lambda \psi _{\lambda },$ \ $%
L_{ext}^{\ast }\ \varphi _{\lambda }=\ \bar{\lambda}\varphi _{\lambda },$ \ $%
(\psi _{\lambda },\varphi _{\lambda })\in \mathcal{H}_{--}\times \mathcal{H}%
_{--}$ and \ $\lambda \in \sigma (L)\ \cap \ \bar{\sigma}(L^{\ast }),$ where 
\begin{equation}
\mathcal{H}_{++}\subset \mathcal{H}_{+}\subset \mathcal{H}\subset \mathcal{H}%
_{-}\subset \mathcal{H}_{--}  \label{eq:15}
\end{equation}%
is some appropriate nucleous rigging extension of the Hilbert space $%
\mathcal{H}.$
\end{theorem}

\begin{proof}
$\vartriangleleft $ \ It is easy to see owing to (\ref{eq:13}) that the
kernel (\ref{eq:14}) satisfies the equation (\ref{eq:12}). Consider now the
Hilbert-Schmidt rigged chain (\ref{eq:15}) and construct a new Hilbert space 
$\mathcal{H}_{\text{K}}$ with the scalar product 
\begin{equation}
(u,v)_{\text{K}}:=(|\mathrm{\hat{K}}|,v\otimes u)_{\mathcal{H}\otimes 
\mathcal{H}}  \label{eq:16}
\end{equation}%
for all $u,v\in \mathcal{H}_{+},$ where $|\mathrm{\hat{K}}|:=\sqrt{\mathrm{%
\hat{K}}^{\ast }\mathrm{\hat{K}}}\in \mathcal{H}_{-}\otimes \mathcal{H}_{-}$
is a positive defined kernel.The norm $||\cdot ||_{\mathrm{K}}$ in $\mathcal{%
H}_{+}$ is, evidently, weaker from the norm $||\cdot ||_{+}$ in $\mathcal{H}%
_{+}$ since for any $u\in \mathcal{H}_{+}$ $\ \ \ |\mathrm{|}u\mathrm{||}%
_{+}^{2}\mathrm{=(|\hat{K}|,}u\otimes u\mathrm{)}_{\mathcal{H}\otimes 
\mathcal{H}}$ $\leq ||\mathrm{\hat{K}}||_{-}||u\otimes u||_{+}$ $=||\mathrm{%
\hat{K}}||_{-}||u||_{+}^{2},$ thereby there holds the embedding $\mathcal{H}%
_{+}\subset \mathcal{H}_{\text{K}}$. Introduce now a related with the
Hilbert space $\mathcal{H}_{\text{K}}$ rigged chain 
\begin{equation}
\mathcal{H}_{++}\subset (\mathcal{H}_{+})\subset \mathcal{H}_{\mathrm{K}%
}\subset \mathcal{H}_{-\text{ }-,\mathrm{K}}.  \label{eq:17}
\end{equation}%
and consider \ (see \cite{Be,BS}) the following expression 
\begin{equation}
(u,\mathrm{E}(\Delta )v)_{K}=\int_{\Delta }\mathrm{d\rho }_{\sigma }(\lambda
)(\mathrm{u},\mathrm{P(\lambda )}v\mathrm{)}_{\mathrm{K}}\mathrm{,}
\label{eq:18}
\end{equation}%
where $\Delta \subset \mathbb{C},$ $u,v\in \mathcal{H}_{++},$ $\mathrm{E:}%
\mathbb{C}\mathrm{\rightarrow }\mathcal{B}\mathrm{(}\mathcal{H}_{++}\QTR{sl}{%
)}$ is a generalized unity spectral expansion for our spectral operator $L:%
\mathcal{H}\rightarrow \mathcal{H}$ whose domain $D(L)$ is reduced to $%
\mathcal{H}_{++}\subset D(L),$ P:$\mathbb{C}\rightarrow \mathcal{B}_{2}(%
\mathcal{H}_{++};\mathcal{H}_{-\text{ }-,\mathrm{K}})$ is a Hilbert-Schmidt
operator of generalized projecting and $\mathrm{\rho }_{\sigma }$ is some
finite Borel measure on $\mathbb{C}$. Assume now that our kernel \^{K}$\in 
\mathcal{H}_{-}\otimes \mathcal{H}_{-}$ allows the following representation:%
\begin{equation}
\mathrm{(\hat{K},}v\mathrm{\otimes u)}_{\mathcal{H}\otimes \mathcal{H}}=%
\mathrm{(|K|,S}_{\mathrm{K}}^{(r)}v\mathrm{\otimes \mathrm{S}_{\mathrm{K}%
}^{(l)}u)}_{\mathcal{H}\otimes \mathcal{H}}=\mathrm{(\mathrm{S}_{\mathrm{K}%
}^{(l)}u,\mathrm{S}_{\mathrm{K}}^{(r)}}v\mathrm{)}_{\mathrm{K}}
\label{eq.19}
\end{equation}%
for all $u\mathrm{,}v\in \mathcal{H}_{++},$ where $\mathrm{S}_{\mathrm{K}%
}^{(l)}$ and $\mathrm{\mathrm{S}_{\mathrm{K}}^{(r)}}\in \mathcal{B}\mathrm{(}%
\mathcal{H}_{++}\QTR{sl}{)}$ are some appropriate bounded operator in $%
\mathcal{H}_{++}.$ Then making use of (\ref{eq:17}) and (\ref{eq:18}) one
finds that 
\begin{equation}
\mathrm{(\hat{K},}v\mathrm{\otimes u)}_{\mathcal{H}\otimes \mathcal{H}%
}=\int_{\sigma (L)\cap \bar{\sigma}(L^{\ast })}\mathrm{d\rho }_{\sigma
}(\lambda )(\mathrm{\mathrm{S}_{\mathrm{K}}^{(l),-1}}u,\mathrm{P(\lambda )%
\mathrm{S}_{\mathrm{K}}^{(r),-1}}v\mathrm{)}_{\mathrm{K}}\mathrm{.}
\label{eq:19.1}
\end{equation}%
Since there exist \cite{Be,BS} two\ isometries $\mathrm{J}_{\mathrm{K}}$ :$%
\mathcal{H}_{-\text{ }-,\mathrm{K}}\rightarrow \mathcal{H}_{++}$ and \textrm{%
J} : $\mathcal{H}_{--}\rightarrow \mathcal{H}_{++}$\ related, respectively,
with rigged chains (\ref{eq:17}) and (\ref{eq:15}), the scalar product on
the right-hand side \ of (\ref{eq:19.1}) can be transformed as 
\begin{equation}
(\mathrm{\mathrm{S}_{\mathrm{K}}^{(l),-1}}u,\mathrm{P(\lambda )\mathrm{S}_{%
\mathrm{K}}^{(r),-1}}v\mathrm{)}_{\mathrm{K}}=(\mathrm{\mathrm{S}_{\mathrm{K}%
}^{(l),-1}}u,\mathrm{\mathrm{J}_{\mathrm{K}}P(\lambda )\mathrm{S}_{\mathrm{K}%
}^{(r),-1}}v\mathrm{)}_{\mathrm{++}}  \label{eq:19.2}
\end{equation}%
for all $u\mathrm{,}v\in \mathcal{H}_{++},$ where $\mathrm{\mathrm{J}_{%
\mathrm{K}}\mathrm{S}_{\mathrm{K}}P(\lambda )\in }\mathcal{B}_{2}\mathrm{(}%
\mathcal{H}_{++}\QTR{sl}{)}$ as a product of the Hilbert-Schmidt operator $%
\mathrm{P(\lambda )}$ $\mathrm{\in }\mathcal{B}_{2}(\mathcal{H}_{++};%
\mathcal{H}_{-\text{ }-,\mathrm{K}}),$ $\lambda \in \sigma (L)\cap \bar{%
\sigma}(L^{\ast }),$ and bounded operators \textrm{J}$_{\mathrm{K}}$ and $%
\mathrm{\mathrm{S}_{\mathrm{K}}.}$ Then owing to a simple corollary from the
Schwartz theorem 2.1 one gets easily that there exists a kernel \textrm{\^{Q}%
}$_{\lambda }\in \mathcal{H}_{++}\otimes \mathcal{H}_{++}$ for any $\lambda
\in \sigma (L)\cap \bar{\sigma}(L^{\ast }),$ such that 
\begin{equation}
\mathrm{(\mathrm{S}_{\mathrm{K}}^{(l),-1}}u,\mathrm{\mathrm{J}_{\mathrm{K}%
}P(\lambda )\mathrm{S}_{\mathrm{K}}^{(r),-1}}v\mathrm{)}_{\mathrm{++}}%
\mathrm{=(\hat{Q}}_{\lambda },v\mathrm{\otimes }u\mathrm{)}_{\mathcal{H}%
\otimes \mathcal{H}}  \label{eq:19.3}
\end{equation}%
for all $u\mathrm{,}v\in \mathcal{H}_{++}.$ Defining now the kernel \textrm{%
\^{Z}}$_{\lambda }\mathrm{:=(J}^{-1}\mathrm{\otimes J}^{-1}\mathrm{)\hat{Q}}%
_{\lambda }\in \mathcal{H}_{--}\otimes \mathcal{H}_{--},$ $\lambda \in
\sigma (L)\cap \bar{\sigma}(L^{\ast }),$ one finds finally from (\ref%
{eq:19.1}) -(\ref{eq:19.3}) that for all $u\mathrm{,}v\in \mathcal{H}_{++}$
\ 
\begin{equation}
\mathrm{(\hat{K},}v\mathrm{\otimes }u\mathrm{)}_{\mathcal{H}\otimes \mathcal{%
H}}=\int_{\sigma (L)\cap \bar{\sigma}(L^{\ast })}\mathrm{d\rho }_{\sigma
}(\lambda )(\mathrm{\hat{Z}}_{\lambda },v\mathrm{\mathrm{\otimes }}u\mathrm{)%
}_{\mathcal{H}\otimes \mathcal{H}}.  \label{eq:19.4}
\end{equation}%
As the constructed kernel \textrm{\^{Z}}$_{\lambda }\mathrm{\in }\mathcal{H}%
_{--}\otimes \mathcal{H}_{--}$ satisfies, evidently, for every $\lambda \in
\sigma (L)\cap \bar{\sigma}(L^{\ast })$ the relationship (\ref{eq:13}), the
theorem is proved.$\triangleright $
\end{proof}

The construction done above for a self-similar congruent kernel $\mathrm{%
\hat{K}}\in \mathcal{H}_{-}\otimes \mathcal{H}_{-}$ in the form (\ref{eq:14}%
) related with a given operator $L\in \mathcal{L(H)}$ appears to be very
inspiring if the condition self-similarity to replace by a simple
similarity. This topic will be in part discussed below.

\section{Congruent kernel operators, related Delsarte transmutation mappings
and their structure}

\setcounter{equation}{0}3.1. Consider in a Hilbert space $\mathcal{H}$ a
pair of densely defined linear differential operators $L$ and $\tilde{L}\in 
\mathcal{L(H)}.$ The following definition will be useful.

Let a pair of kernels $\mathrm{\hat{K}}_{\mathrm{s}}\in \mathcal{H}%
_{-}\otimes \mathcal{H}_{-},$ $s=\pm ,$ satisfy the following congruence
relationships 
\begin{equation}
(\tilde{L}_{ext}\otimes \mathrm{1})\mathrm{\hat{K}}_{\mathrm{s}}=(\mathrm{1}%
\otimes L_{ext}^{\ast })\mathrm{\hat{K}}_{\mathrm{s}}  \label{eq:20}
\end{equation}%
for a given pair of the respectively extended linear operators $L,\ \tilde{L}%
\in \mathcal{L(H)}.$ Then the kernels $\mathrm{\hat{K}}_{\mathrm{s}}\in 
\mathcal{H}_{-}\otimes \mathcal{H}_{-},$ $s=\pm ,$ will be called congruent
to this pair $(L,\tilde{L})$ of operators in $\mathcal{H}.$

Since not any pair of operators $L,\ \tilde{L}\in \mathcal{L(H)}$ can be
congruent, the natural problem arises if they exist: how to describe the set
of corresponding kernels $\mathrm{\hat{K}}_{\mathrm{s}}\in \mathcal{H}%
_{-}\otimes \mathcal{H}_{-}$ $,$ $s=\pm ,$ congruent to a given pair $(L,%
\tilde{L})$ of operators in $\mathcal{H}.$ The first question being
important for further is that of existence of kernels $\mathrm{\hat{K}}_{%
\mathrm{s}}\in \mathcal{H}_{-}\otimes \mathcal{H}_{-},$ $s=\pm ,$ congruent
to this pair. The question has an evident answer for the case when $\tilde{L}%
=L$ and the congruence is then self- similar. The interesting case when $%
\tilde{L}\neq L$ appears to be very nontrivial and can be treated more or
less successfully if there exist such bounded and invertible operators 
\textbf{$\Omega $}$_{\mathrm{s}}\in \mathcal{B}(\mathcal{H)},$ $s=\pm ,$
that the transmutation conditions 
\begin{equation}
\tilde{L}\mathbf{\Omega }_{\mathrm{s}}=\mathbf{\Omega }_{\mathrm{s}}L
\label{eq:21}
\end{equation}%
hold.

\begin{definition}
(Delsarte, Lions; \cite{De,DL}) Let a pair of densely defined differential
closeable operators $L,\tilde{L}\in \mathcal{L(H)}$ in a Hilbert space $%
\mathcal{H}$ is endowed with a pair of closed subspaces $\mathcal{H}_{0},%
\mathcal{\tilde{H}}_{0}\subset \mathcal{H}_{-}$ subject to a rigged Hilbert
spaces chain (\ref{0}). Then invertible operators \textbf{$\Omega $}$_{%
\mathrm{s}}\in Aut(\mathcal{H})\cap \mathcal{B(H)},$ $s=\pm ,$ are called 
\textbf{Delsarte transmutations} if the following conditions hold:

\begin{itemize}
\item[i)] the operators $\mathbf{\Omega }_{\mathrm{s}}$ and their inverse $%
\mathbf{\Omega }_{\mathrm{s}}^{-1},$ $s=\pm ,$ are continuous in $\mathcal{H}%
,$ that is \textbf{$\Omega $}$_{\mathrm{s}}\in Aut(\mathcal{H})\cap \mathcal{%
B}(\mathcal{H}),s=\pm ;$

\item[ii)] the images $\mathrm{Im}\ $\textbf{$\Omega $}$_{\mathrm{s}}|_{%
\mathcal{H}_{0}}=\mathcal{\tilde{H}}_{0},s=\pm ;$

\item[iii)] the relationships (\ref{eq:21}) are satisfied.
\end{itemize}
\end{definition}

Suppose now that an operator pair $(L,\tilde{L})$ $\subset \mathcal{L}(%
\mathcal{H)}$ is differential of the same order $n(L)\in \mathbb{Z}_{+},$
that is the following representations 
\begin{equation}
L:=\sum_{|\alpha |=0}^{n(L)}a_{\alpha }(x)\frac{\partial ^{|\alpha |}}{%
\partial x^{\alpha }},\ \ \ \ \tilde{L}:=\sum_{|\alpha |=0}^{n(L)}\tilde{a}%
_{\alpha }(x)\frac{\partial ^{|\alpha |}}{\partial x^{\alpha }},
\label{eq:22}
\end{equation}%
hold, where $x\in \mathrm{Q},$ $\mathrm{Q}\subset \mathbb{R}^{m}$ is some
open connected region in $\mathbb{R}^{m},$ the smooth coefficients $%
a_{\alpha },$ $\tilde{a}_{\alpha }$ $\in C^{\infty }(\mathrm{Q};End\ \mathbb{%
C}^{\mathrm{N}})$ for all $\alpha \in \mathbb{Z}_{+}^{m},$ $|\alpha |=%
\overline{0,n(L)}$ and $\mathrm{N}\in \mathbb{Z}_{+}.$ The differential
expressions (\ref{eq:22}) are defined and closeable on the dense in the
Hilbert space $\mathcal{H}:=L_{2}(\mathrm{Q};\mathbb{C}^{\mathrm{N}})$
domains $\mathrm{D}(L),$ $\mathrm{D}(\tilde{L})$ $\subset $ $\mathrm{W}%
_{2}^{n(L)}(\mathrm{Q};\mathbb{C}^{\mathrm{N}})\subset \mathcal{H}.$ This,
in particular, means that there exists the corresponding to (\ref{eq:22})
pair of adjoint operators $L^{\ast },\tilde{L}^{\ast }\in \mathcal{L}(%
\mathcal{H})$ which are defined also on dense domains $\mathrm{D}(L^{\ast
}), $ $\mathrm{D}(\tilde{L}^{\ast })\subset \mathrm{W}_{2}^{n(L)}(\mathrm{Q};%
\mathbb{C}^{\mathrm{N}})\subset \mathcal{H}.$

Take now a pair of invertible bounded linear operators $\ \mathbf{\Omega }_{%
\mathrm{s}}\in \mathcal{L(H)},$ $s=\pm ,$ and look at the following Delsarte
transformed operators 
\begin{equation}
\tilde{L}_{\mathrm{s}}:=\mathbf{\Omega }_{\mathrm{s}}L\mathbf{\Omega }_{%
\mathrm{s}}^{-1},\text{ }  \label{eq:23}
\end{equation}%
\ $s=\pm ,$ which, by definition, must persist to be also differential. An
additional natural constraint involved on operators \textbf{$\Omega $}$%
_{s}\in Aut(\mathcal{H})\cap \mathcal{B(H)},$ $s=\pm ,$ is the independence 
\cite{Fa,FT} of differential expressions for operators (\ref{eq:23}) on
indices $s=\pm .$ The problem of constructing such Delsarte transmutation
operators \textbf{$\Omega $}$_{s}\in Aut(\mathcal{H})\cap \mathcal{B(H)},$ $%
s=\pm ,$ appeared to be very complicated and in the same time dramatic as it
one could observe from special results obtained in \cite{Fa,Ni} for
two-dimensional Dirac and three-dimensional Laplace type operators.

3.2. Before proceeding to setting up our approach to treating the problem
mentioned above, let us consider some formal generalizations of the results
described before in Section 2. Take an elementary kernel $\widehat{\tilde{Z}}%
_{\lambda }\in \mathcal{H}_{-}\otimes \mathcal{H}_{-}$ satisfying the
conditions generalizing (\ref{eq:11}): 
\begin{equation}
(\tilde{L}_{ext}\otimes \mathrm{I})\ \widehat{\mathrm{\tilde{Z}}}_{\lambda
}=\lambda \widehat{\mathrm{\tilde{Z}}}_{\lambda },\text{ \ \ \ \ }(\mathrm{I}%
\otimes L_{ext}^{\ast })\ \widehat{\mathrm{\tilde{Z}}}_{\lambda }=\lambda 
\widehat{\mathrm{\tilde{Z}}}_{\lambda }  \label{eq:24}
\end{equation}%
for $\lambda \in \sigma (\tilde{L})\cap \bar{\sigma}(L^{\ast }),$ being,
evidently, well suitable for treating the equation (\ref{eq:20}). Then one
sees that an elementary kernel $\widehat{\mathrm{Z}}_{\lambda }\in \mathcal{H%
}_{-}\otimes \mathcal{H}_{-}$ for any $\lambda \in \sigma (\widetilde{L}%
)\cap \sigma (\widetilde{L}^{\ast })$ solves the equation (\ref{eq:20}),
that is 
\begin{equation}
(\widetilde{L}_{ext}\otimes \mathrm{1})\ \widehat{\mathrm{\tilde{Z}}}%
_{\lambda }=(\mathrm{1}\otimes L_{ext}^{\ast })\ \widehat{\mathrm{\tilde{Z}}}%
_{\lambda }.  \label{eq:25}
\end{equation}%
Thereby one can expect that for kernels $\mathrm{\hat{K}}_{\mathrm{s}}\in 
\mathcal{H}_{-}\otimes \mathcal{H}_{-},$ $s=\pm ,$ there exist the similar
to (\ref{eq:16}) spectral representations 
\begin{equation}
\mathrm{\hat{K}}_{\mathrm{s}}=\int_{\sigma (\widetilde{L})\cap \bar{\sigma}%
(L^{\ast })}\widehat{\mathrm{\tilde{Z}}}_{\lambda }\mathrm{d}\rho _{\sigma ,%
\mathrm{s}}(\lambda ),  \label{eq:26}
\end{equation}%
$s=\pm ,$ with finite spectral measures $\rho _{\sigma ,\mathrm{s}},$ $s=\pm
,$ localized upon the Borel subsets of the common spectrum $\sigma (\tilde{L}%
)\cap \bar{\sigma}(L^{\ast }).$ Based of the spectral representation like (%
\ref{eq:15}) applied separately to operators $\tilde{L}\in \mathcal{L}(%
\mathcal{H})$ and $L^{\ast }\in \mathcal{L}(\mathcal{H})$ one states
similarly as before the following theorem.

\begin{theorem}
The equations (\ref{eq:24}) are compatible for any $\lambda \in \sigma (%
\widetilde{L})\cap \bar{\sigma}(L^{\ast })$ and, moreover, for kernels $%
\mathrm{\hat{K}}_{\mathrm{s}}\in \mathcal{H}_{-}\otimes \mathcal{H}_{-},$ $%
s=\pm ,$ \ satisfying the congruence condition (\ref{eq:20}) there exist a
kernel $\widehat{\mathrm{\tilde{Z}}}_{\lambda }\in $ $\mathcal{H}_{-\text{ }%
-}\otimes \mathcal{H}_{-\text{ }-}$ \ for a suitably Gelfand rigged Hilbert
spaces chain (\ref{eq:18}), such that the spectral representations (\ref%
{eq:26}) hold.
\end{theorem}

Now we will be interested in the inverse problem of constructing kernels $%
\mathrm{\hat{K}}_{s}\in \mathcal{H}_{-}\otimes \mathcal{H}_{-},$ $s=\pm ,$
like (\ref{eq:26}) a priori satisfying the congruence conditions (\ref{eq:20}%
) subject to the same pair $(L,\tilde{L})$ of differential operators in $%
\mathcal{H}$ and related via the Delsarte transmutation condition (\ref%
{eq:21}). In some sense we shall state that only for such Delsarte related
operator pairs $(L,\tilde{L})$ in $\mathcal{H}$ one can construct a dual
pair $\{\mathrm{\hat{K}}_{s}\in $ $\mathcal{H}_{-}\otimes \mathcal{H}_{-}$ $%
:s=\pm \}$ of the corresponding congruent kernels satisfying the conditions
like (\ref{eq:20}), that is \ 
\begin{equation}
(\tilde{L}_{ext}\otimes \mathrm{1})\ \mathrm{\hat{K}}_{\pm }=\mathrm{\hat{K}}%
_{\pm }(\mathrm{1}\otimes L_{ext}^{\ast }).  \label{eq:26a}
\end{equation}

3.3. Suppose now that there exists another pair of Delsarte transmutation
operators $\mathbf{\Omega }_{s}$ and $\mathbf{\Omega }_{s}^{\circledast }\
\in $ $Aut(H)$ $\cap \mathcal{B(H)},$ $s=\pm ,$ satisfying condition ii) of
Definition 3.2 subject to the corresponding two pairs of differential
operators $(L,\tilde{L})$ and $(L^{\ast },\tilde{L}^{\ast })$ $\subset 
\mathcal{L(H)}.$ This means, in particular, that there exists an additional
pair of closed subspace $\mathcal{H}_{0}^{\circledast }$ and $\mathcal{%
\tilde{H}}_{0}^{\circledast }\subset \mathcal{H}_{-}$ such that 
\begin{equation}
\mathrm{Im}\ \mathbf{\Omega }_{s}^{\circledast }|_{\mathcal{H}%
_{0}^{\circledast }}=\mathcal{\tilde{H}}_{0}^{\circledast }  \label{eq:27}
\end{equation}%
$s=\pm ,$ for the Delsarte transmutation operator $\mathbf{\Omega }%
_{s}^{\circledast }\in Aut(H)\cap \mathcal{B(H)},$ $s=\pm ,$ satisfying the
obvious conditions 
\begin{equation}
\tilde{L}^{\ast }\cdot \mathbf{\Omega }_{s}^{\circledast }=\mathbf{\Omega }%
_{s}^{\circledast }\cdot L^{\ast }  \label{eq:28}
\end{equation}%
$s=\pm ,$ involving the adjoint operators $\tilde{L}^{\ast },L^{\ast }\in 
\mathcal{L}(\mathcal{H})$ defined before and given by the following from (%
\ref{eq:22}) usual differential expressions: 
\begin{equation}
L^{\ast }=\sum_{|\alpha |=0}^{n(L)}(-1)^{|\alpha |}\frac{\partial ^{|\alpha
|}}{\partial x^{\alpha }}\cdot \bar{a}_{\alpha }^{\mathrm{\intercal }}(x),%
\text{ }\tilde{L}^{\ast }=\sum_{|\alpha |=0}^{n(L)}(-1)^{|\alpha |}\frac{%
\partial ^{|\alpha |}}{\partial x^{\alpha }}\cdot \overset{\_}{\tilde{a}}%
_{\alpha }^{\mathrm{\intercal }}(x)  \label{eq:29}
\end{equation}%
for all $x\in \mathrm{Q}\subset \mathbb{R}^{m}.$

Construct now the following \cite{GK,My} Delsarte transmutation operators of
Volterra type 
\begin{equation}
\mathbf{\Omega }_{\pm }:=\mathrm{1}\ +\mathrm{K}_{\pm }(\mathbf{\Omega }),
\label{eq:30}
\end{equation}%
corresponding to some two different kernels $\mathrm{\hat{K}}_{+}$ and $%
\mathrm{\hat{K}}_{-}\in \mathcal{H}_{-}\otimes \mathcal{H}_{-},$ of integral
Volterrian operators $\mathrm{K}_{+}(\mathbf{\Omega })$ and $\mathrm{K}_{-}(%
\mathbf{\Omega })$ related with them in the following way: 
\begin{equation}
(u,\mathrm{K}_{\pm }(\mathbf{\Omega })v):=(u\ \chi (S_{x,\pm }^{(m)}),%
\mathrm{\hat{K}}_{\pm }v)  \label{eq:31}
\end{equation}%
for all $(u,v)\in \mathcal{H}_{+}\times \mathcal{H}_{+},$ where $\chi
(S_{x,\pm }^{(m)})$ are some characteristic functions of two $m$-dimensional
smooth hypersurfaces $S_{x,+}^{(m)}$ and $S_{x,-}^{(m)}\in \mathcal{K}(%
\mathrm{Q})$ from a singular simplicial complex $\mathcal{K}(\mathrm{Q})$ of
the open set $\mathrm{Q}\subset \mathbb{R}^{m},$ chosen such that the
boundary $\partial (S_{x,+}^{(m)}\cup S_{x,-}^{(m)})=\partial \mathrm{Q}.$
In the case when $Q:=\mathbb{R}^{m},$ it is assumed naturally that $\partial 
\mathbb{R}^{m}=\oslash .$ Making use of the Delsarte operators (\ref{eq:30})
and relationship like (\ref{eq:21}) one can construct the following
differential operator expressions: 
\begin{equation}
\tilde{L}_{\pm }-L=\mathrm{K}_{\pm }(\mathbf{\Omega })L-\tilde{L}_{\pm }%
\mathrm{K}_{\pm }(\mathbf{\Omega }).  \label{eq:32}
\end{equation}%
Since the left-hand sides of (\ref{eq:32}) are, by definition, purely
differential expressions, one follows right away that the local kernel
relationships like (\ref{eq:26}) hold: 
\begin{equation}
(\tilde{L}_{ext,\pm }\otimes \mathrm{1})\ \mathrm{\hat{K}}_{\pm }=(\mathrm{1}%
\otimes L_{ext}^{\ast })\ \mathrm{\hat{K}}_{\pm }.  \label{eq:33}
\end{equation}%
The expressions (\ref{eq:32}) define, in general, two different differential
expressions $\tilde{L}_{\pm }\in \mathcal{L}(\mathcal{H})$ depending
respectively both on the kernels $\mathrm{\hat{K}}_{\pm }\in \mathcal{H}%
_{-}\otimes \mathcal{H}_{-}$ and on the chosen hypersurfaces $S_{x,\pm
}^{(m)}\in \mathcal{K}(\mathrm{Q}).$ As will be stated later, the following
important theorem holds.

\begin{theorem}
Let smooth hypersurfaces $S_{x,\pm }^{(m)}\in \mathcal{K}(\mathrm{Q})$ be
chosen in such a way that $\partial (S_{x,+}^{(m)}\cup
S_{x,-}^{(m)})=\partial \mathrm{Q}$ and $\partial S_{x,\pm }^{(m)}=\mp
\sigma _{x}^{(m-1)}+\sigma _{x_{\pm }}^{(m-1)},$ where $\sigma _{x}^{(m-1)}$
and $\sigma _{x_{\pm }}^{(m-1)}$ are some homological subject to the the
homology group $\mathrm{H}_{m-1}(\mathrm{Q};\mathbb{C})$ simplicial chains,
parametrized, respectively, by a running point $x\in \mathrm{Q}$ and fixed
points $x_{\pm }\in \partial \mathrm{Q}$ and satisfying the following
homotopy condition: $\lim_{x\rightarrow x_{\pm }}$ $\ \sigma _{x}^{(m-1)}$ $%
=\mp \sigma _{x_{\pm }}.$ Then the operator equalities 
\begin{equation}
\tilde{L}_{+}:=\mathbf{\Omega }_{+}L\mathbf{\Omega }_{+}^{-1}=\tilde{L}=%
\mathbf{\Omega }_{-}L\mathbf{\Omega }_{-}^{-1}:=\tilde{L}_{-}  \label{eq:34}
\end{equation}%
are satisfied if the following commutation property 
\begin{equation}
\lbrack \mathbf{\Omega }_{+}^{-1}\ \mathbf{\Omega }_{-},L]=0  \label{eq:34a}
\end{equation}%
or, equivalently, kernel relationship 
\begin{equation}
(L_{ext}\otimes \mathrm{1})\mathbf{\hat{\Omega}}_{+}^{-1}\ast \mathbf{\hat{%
\Omega}}_{-}=(\mathrm{1}\otimes L_{ext}^{\ast })\mathbf{\hat{\Omega}}%
_{+}^{-1}\ast \mathbf{\hat{\Omega}}_{-}  \label{eq:34b}
\end{equation}%
hold.

\begin{remark}
It is a place to notice here that special degenerate cases of theorem 3.4 \
where before proved in works \cite{Fa,Ni} \ for two-dimensional Dirac and
three-dimensional Laplace type differential operators. The constructions and
tools devised \ there appeared to be instructive and motivative for the
approach developed here by us in the general case.
\end{remark}
\end{theorem}

3.4. Consider now a pair $(\mathbf{\Omega }_{+},\mathbf{\Omega }_{-})$ of
Delsarte transmutation operators being in the form (\ref{eq:30}) and
respecting all of the conditions from Theorem 3.4. Then the following lemma
is true.

\begin{lemma}
Let an invertible Fredholm operator $\mathbf{\Omega }=1+\Phi (\mathbf{\Omega 
})\in Aut(\mathcal{H})\cap \mathcal{B(H)}$ with $\Phi \in \mathcal{B}%
_{\infty }(\mathcal{H})$ allow the factorization representation 
\begin{equation}
\mathbf{\Omega }=\mathbf{\Omega }_{+}^{-1}\mathbf{\Omega }_{-}  \label{eq:35}
\end{equation}%
by means of two Delsarte operators $\mathbf{\Omega }_{+}$ and $\mathbf{%
\Omega }_{-}\in Aut(\mathcal{H})\cap \mathcal{B(H)}$ in the form (\ref{eq:30}%
). Then there exists the unique operator kernel $\hat{\Phi}\in \mathcal{H}%
_{-}\otimes \mathcal{H}_{-}$ corresponding naturally to the compact operator 
$\Phi (\mathbf{\Omega })\in \mathcal{B}_{\infty }(\mathcal{H})$ and
satisfying the following self-similar congruence commutation condition: 
\begin{equation}
(L_{ext}\otimes \mathrm{1})\ \hat{\Phi}=(\mathrm{1}\otimes L_{ext}^{\ast })\ 
\hat{\Phi},  \label{eq:36}
\end{equation}%
related to the properties (\ref{eq:34a} and (\ref{eq:34b}).
\end{lemma}

From the equality (\ref{eq:36}) and Theorem 2.2 one gets easily the
following corollary. 
\begin{corollary}
There exists such a finite Borel measure $\rho_{\sigma}$ defined on the Borel
subsets of $\sigma(\mathrm{L}) \cap \overline{\sigma}(\mathrm{L}%
^{*})$, that the following weak equality
\begin{equation}
\widehat{\Phi}= \int_{\sigma(\mathrm{L}) \cap \overline{\sigma}(\mathrm{L}%
^{*})} \widehat{\mathrm{Z}}_{\lambda} \mathrm{d} \rho_{\sigma}(\lambda)
\label{eq:37}
\end{equation}
holds.
\end{corollary} \ 

Concerning the differential expression $L\in \mathcal{L}(\mathcal{H})$ and
the corresponding Volterra type Delsarte transmutation operators $\mathbf{%
\Omega }_{\pm }\in \mathcal{B}_{\infty }(\mathcal{H})$ the conditions (\ref%
{eq:34a}) and (\ref{eq:36}) are equivalent to the operator equation

\begin{equation}
\lbrack \Phi (\mathbf{\Omega }),L]=0.  \label{eq:38}
\end{equation}%
Really, since equalities (\ref{eq:34}) hold, one gets easily that

\begin{equation*}
L(\mathrm{1}+\Phi (\mathbf{\Omega }))=L(\mathbf{\Omega }_{+}^{-1}\mathbf{%
\Omega }_{-})=\mathbf{\Omega }_{+}^{-1}(\mathbf{\Omega }_{+}L\mathbf{\Omega }%
_{+}^{-1})\mathbf{\Omega }_{-}
\end{equation*}%
\begin{equation}
=\mathbf{\Omega }_{+}^{-1}(\mathbf{\Omega }_{-}L\mathbf{\Omega }_{-})\mathbf{%
\Omega }_{-}=\mathbf{\Omega }_{+}^{-1}\mathbf{\Omega }_{-}L=(\mathrm{1}+\Phi
(\mathbf{\Omega }))L,  \label{eq:39}
\end{equation}%
meaning exactly (\ref{eq:38}).

Suppose also that, first, for another Fredholm operator $\mathbf{\Omega }%
^{\circledast }=\mathrm{1}+\Phi ^{\circledast }(\mathbf{\Omega })\in Aut(%
\mathcal{H})\cap \mathcal{B(H)}$ with $\Phi ^{\circledast }(\mathbf{\Omega }%
)\in \mathcal{B}_{\infty }(\mathcal{H})$ there exist two factorizing it
Delsarte transmutation Volterra type operators $\mathbf{\Omega }_{\pm
}^{\circledast }\in Aut(\mathcal{H})\cap \mathcal{B(H)}$ in the form 
\begin{equation}
\mathbf{\Omega }_{\pm }^{\circledast }=\mathrm{1}+\mathrm{K}_{\pm
}^{\circledast }(\mathbf{\Omega })  \label{eq:40}
\end{equation}%
with Volterrian \cite{GK} integral operators $\mathrm{K}_{\pm }^{\circledast
}(\mathbf{\Omega })$ related naturally with some kernels $\mathrm{\hat{K}}%
_{\pm }\in \mathcal{H}_{-}\otimes \mathcal{H}_{-},$ and, second, the
factorization condition 
\begin{equation}
\mathrm{1}+\Phi ^{\circledast }(\mathbf{\Omega })=\mathbf{\Omega }%
_{+}^{\circledast ,-1}\mathbf{\Omega }_{-}^{\circledast }  \label{eq:41}
\end{equation}%
is satisfied, then the following theorem holds.

\begin{theorem}
Let a pair of hypersurfaces $S_{x,\pm }^{(m)}\subset \mathcal{K}(\mathrm{Q})$
satisfy all of the conditions from Theorem 3.2. Then the Delsarte
transformed operators $\tilde{L}_{\pm }^{\ast }\in \mathcal{L}(\mathcal{H})$
are differential and equal, that is 
\begin{equation}
\tilde{L}_{+}^{\ast }=\mathbf{\Omega }_{+}^{\circledast }L^{\ast }\mathbf{%
\Omega }_{+}^{\circledast ,-1}=\tilde{L}^{\ast }=\mathbf{\Omega }%
_{-}^{\circledast }L^{\ast }\mathbf{\Omega }_{-}^{\circledast ,-1}=\tilde{L}%
_{-}^{\circledast },  \label{eq:42}
\end{equation}%
iff the following commutation condition 
\begin{equation}
\lbrack \Phi ^{\circledast }(\mathbf{\Omega }),L^{\ast }]=0  \label{eq:43}
\end{equation}%
holds.
\end{theorem}

\begin{proof}
$\vartriangleleft $ \ A proof of this theorem is stated by reasonings
similar to those done before when analyzing the congruence condition for a
given pair $(L,\tilde{L})\subset \mathcal{L(H)}$ of differential operators
and their adjoint ones in $\mathcal{H}.\vartriangleright $
\end{proof}

By means of the Delsarte transmutation from the differential operators $L$
and $L^{\ast }\in \mathcal{L}(\mathcal{H})$ we have obtained above two
differential operators 
\begin{equation}
\tilde{L}=\mathbf{\Omega }_{\pm }L\mathbf{\Omega }_{\pm }^{-1},\ \ \tilde{L}%
^{\ast }=\mathbf{\Omega }_{\pm }^{\circledast }L^{\ast }\mathbf{\Omega }%
_{\pm }^{\circledast ,-1},  \label{eq:44}
\end{equation}%
which must be compatible and, thereby, related as 
\begin{equation}
(\tilde{L})^{\ast }=\widetilde{(L^{\ast })}.  \label{eq:45}
\end{equation}%
The condition (\ref{eq:45}) due to (\ref{eq:44}) gives rise to the following
additional commutation expressions for kernels $\mathbf{\Omega }_{\pm
}^{\circledast }$ and$\ \ \mathbf{\Omega }_{\pm }^{\ast }\in Aut(\mathcal{H}%
)\cap \mathcal{B}(\mathcal{H)}:$ 
\begin{equation}
\lbrack L^{\ast },\mathbf{\Omega }_{\pm }^{\ast }\mathbf{\Omega }_{\pm
}^{\circledast }]=0,  \label{eq:46}
\end{equation}%
being equivalent, obviously, to such a commutation relationship: 
\begin{equation}
\lbrack L,\mathbf{\Omega }_{\pm }^{\circledast ,\ast }\mathbf{\Omega }_{\pm
}]=0.  \label{eq:47}
\end{equation}%
As a result of representations (\ref{eq:47}) one can formulate the following
corollary.

\begin{corollary}
There exist finite Borel measures $\rho _{\sigma ,\pm }$ \ localized upon
the common spectrum $\sigma (L)\cap \bar{\sigma}(L^{\ast }),$ such that the
following weak kernel representations
\begin{equation}
\mathbf{\hat{\Omega}}_{\pm }^{\circledast ,\ast }\ast \mathbf{\hat{\Omega}}%
_{\pm }=\int_{\sigma (L)\cap \bar{\sigma}(L^{\ast })}\mathrm{\hat{Z}}%
_{\lambda }\mathrm{d}\rho _{\sigma ,\pm }(\lambda )  \label{eq:48}
\end{equation}%
hold, where $\mathbf{\hat{\Omega}}_{\pm }^{\circledast ,\ast }$ and $\
\mathbf{\hat{\Omega}}_{\pm }\in \mathcal{H}_{-}\otimes \mathcal{H}_{-}$ are
the corresponding kernels of integral Volterrian operators $\mathbf{\Omega }%
_{\pm }^{\circledast ,\ast }$ and $\ \mathbf{\Omega }_{\pm }\in Aut(\mathcal{%
H})\cap \mathcal{B(H)}.$
\end{corollary}

3.5. The integral operators of Volterra type (\ref{eq:30}) constructed above
by means of kernels in the form (\ref{eq:26}) are, as well known \cite%
{LS,Le,Fa,Ma,Bu}, very important for studying many problems of spectral
analysis and related integrable nonlinear dynamical systems \cite%
{FT,Ma,No,Ni,PM} on functional manifolds. \ In particular, they serve as
factorizing operators for a class of Fredholm operators entering the
fundamental Gelfand - Levitan - Marchenko operator equations \cite{LS, Ma,FT}
whose solutions are exactly kernels of Delsarte transmutation operators of
Volterra type, related with the corresponding congruent kernels subject to
given pairs of closeable differential operators in a Hilbert space $\mathcal{%
H}.$ Thereby it is natural to try to learn more of their structure
properties subject to their representations both in the form (\ref{eq:26}), (%
\ref{eq:30}), and in the dual form within the general Gokhberg - Krein
theory \cite{GK,FT,My} of Volterra type operators.

To proceed further with we need to introduce some additional notions and
definitions from \cite{GK,Bu} important for what will follow below. \ Define
a set $\mathcal{P}$ of projectors $\mathrm{P}^{2}=\mathrm{P}:\mathcal{H}\
\rightarrow \mathcal{H}$ which is called a \textbf{projector chain} if for
any pair $\mathrm{P}_{1},\mathrm{P}_{2}\in \mathcal{P},$ $\mathrm{P}_{1}\neq 
\mathrm{P}_{2},$ one has either $\mathrm{P}_{1}<\mathrm{P}_{2}$ or $\mathrm{P%
}_{2}<\mathrm{P}_{1}$, and $\mathrm{P}_{1}\mathrm{P}_{2}=\min (\mathrm{P}%
_{1},\mathrm{P}_{2}).$ The ordering $\mathrm{P}_{1}<\mathrm{P}_{2}$ above
means, as usually, that $\mathrm{P}_{1}\mathcal{H}\subset \mathrm{P}_{2}%
\mathcal{H}$, $\mathrm{P}_{1}\mathcal{H}\neq \mathrm{P}_{2}\mathcal{H}$. If $%
\mathrm{P}_{1}\mathcal{H}\subset \mathrm{P}_{2}\mathcal{H}$, then one writes
down that $\mathrm{P}_{1}\leq \mathrm{P}_{2}$. The closure $\overline{%
\mathcal{P}}$ of a chain $\mathcal{P}$ means, by definition, that set of all
operators being weak limits of sequences from $\mathcal{P}$. The inclusion
relationship $\mathcal{P}_{1}\subset \mathcal{P}_{2}$ of any two sets of
projector chains possesses obviously the transitivity property allowing to
consider the set of all projector chains as a partly ordered set. A chain $%
\mathcal{P}$ \ is called \textbf{maximal} if it can not be extended. It is
evident that a maximal chain is closed and contains zero $0\in \mathcal{P}$
and unity $\mathrm{1}\in \mathcal{P}$ operators. A pair of projectors $(%
\mathrm{P}^{-},\mathrm{P}^{+})\subset \mathcal{P}$ is called a \textbf{break}
of the chain $\mathcal{P}$ if $\mathrm{P}_{-}<\mathrm{P}_{+}$ and for all $%
\mathrm{P}\in \mathcal{P}$ either $\mathrm{P}<\mathrm{P}^{-}$ or $\mathrm{P}%
^{+}<\mathrm{P}$. A closed chain is called \textbf{continuous} if for any
pair of projectors $\mathrm{P}_{1},\mathrm{P}_{2}\subset \mathcal{P}$ there
exist a projector $\mathrm{P}\in \mathcal{P}$, such that $\mathrm{P}_{1}<%
\mathrm{P}<\mathrm{P}_{2}$. A maximal chain $\mathcal{P}$ will be called
complete if it is continuous. A strongly ascending with to inclusion
projector valued function $\mathrm{P}:\mathrm{Q}\ni \Delta \ \rightarrow \ 
\mathcal{P}$ is called a \textbf{parametrization} of a chain $\mathcal{P}$,
if the chain $\mathcal{P}=Im(\mathbb{P})$ such a parametrization of the
self-adjoint chain $\mathcal{P}$ is called \textbf{smooth}, if for any $u\in 
\mathcal{H}$ the positive value measure $\Delta \ \rightarrow \ (u,\mathrm{P}%
(\Delta )u)$ is absolutely continuous. It is well known \cite{GK,My,FT} that
very complete projector chain allows a smooth parametrization. In what will
follow a projector chain $\mathcal{P}$ will be self-adjoint, complete and
endowed with a fixed smooth parametrization with respect to an operator
valued function $\mathrm{F}:\mathcal{P}\ \rightarrow \ \mathcal{B}(\mathcal{H%
})$ the expressions like $\ \ \int_{\mathcal{P}}\mathrm{F}(\mathrm{P})%
\mathrm{dP}$ \ and $\int_{\mathcal{P}}\mathrm{dPF}(\mathrm{P})$ will be used
for the corresponding \cite{GK} Riemann-Stiltjes integrals subject to the
corresponding projector chain. \ Take now a linear\ compact operator $%
\mathrm{K\in }\mathcal{B}_{\infty }\mathcal{(H)}$ acting in a separable
Hilbert space $\mathcal{H}$ \ endowed with a projector chain $\mathcal{P}.$
A chain $\mathcal{P}$ is also called \textbf{proper} subject to an operator $%
\mathrm{K}\in \mathcal{B}_{\infty }(\mathcal{H})$ if $\mathrm{P}\mathrm{K}%
\mathrm{P}=\mathrm{K}\mathrm{P}$ for any projector $\mathrm{P}\in \mathcal{P}%
,$ meaning obviously that subspace $\mathrm{P}\mathcal{H}$ is invariant with
respect to the operator $\mathrm{K}=\mathcal{B}_{\infty }(\mathcal{H})$ for
any $\mathrm{P}\in \mathcal{P}.$ As before the note by $\sigma (\mathrm{K}(%
\mathbf{\Omega })).$ The spectrum of any operator $\mathrm{K}\in \mathcal{L}(%
\mathcal{H}).$

\begin{definition}
An operator $\mathrm{K}\in \mathcal{B}_{\infty }(\mathcal{H})$ is called 
\textbf{Volterrian} if $\sigma (\mathrm{K})=\{0\}.$
\end{definition}

As it can be shown \cite{GK}, a Volterrian operator $\mathrm{K}\in \mathcal{B%
}_{\infty }(\mathcal{H})$ possesses the maximal proper projector chain $%
\mathcal{P}$ \ such, that for any its break $(\mathrm{P}^{-},\mathrm{P}^{-})$
the following relationship 
\begin{equation}
(\mathrm{P}^{+}-\mathrm{P}^{-})\ \mathrm{K}\ (\mathrm{P}^{+}-\mathrm{P}%
^{-})=0  \label{eq:49}
\end{equation}%
holds. Since integral operators (\ref{eq:30}) constructed before are of
Volterra type and congruent to a pair $(L,\widetilde{L})$ of closeable
differential operators in $\mathcal{H},$ we will be now interested in their
properties with respect both to the definition given above and to the
corresponding proper maximal projector chains $\mathcal{P}(\mathbf{\Omega }%
). $

3.6. Suppose now that we are given a Fredholm operator $\mathbf{\Omega }\in 
\mathcal{B(H)\cap }Aut\mathcal{(H)}$ self- congruent to\ a closeable
differential operator $L\in \mathcal{L(H)}.$ As we are also given with an
elementary kernel (\ref{eq:13}) in the spectral form (\ref{eq:14}), our
present task will be a description of elementary kernels $\mathrm{\hat{Z}}%
_{\lambda },\ $\ $\lambda \in $ $\sigma (L)$ $\cap $ $\bar{\sigma}(L^{\ast
}),$ by means of some smooth and complete parametrization suitable for them.
\ For treating this problem we will make use of very interesting recent
results obtained in \cite{My} and devoted to the factorization problem of
Fredholm operators. As a partial case this work contains some aspects of our
factorization problem for Delsarte transmutation operators $\mathbf{\Omega }%
\in Aut(\mathcal{H)\cap B}(\mathcal{H})$ in the form (\ref{eq:30}).

Let us formulate now some preliminary results from \cite{GK,My} suitable for
the problem under regard. As before, we will the note by $\mathcal{B}(%
\mathcal{H})$ the Banach algebra of all linear and continuous every where
defined operators in $\mathcal{H},$ and also by $\mathcal{B}_{\infty }(%
\mathcal{H})$ the Banach algebra of all compact operators from $\mathcal{B}(%
\mathcal{H})$ and by $\mathcal{B}_{0}(\mathcal{H})$ the linear subspace of
all finite dimensional operator from $\mathcal{B}_{\infty }(\mathcal{H}).$ 
\newline
Put also, by definition, 
\begin{equation}
\mathcal{B}^{-}(\mathcal{H})=\{\mathrm{K}\in \mathcal{B}(\mathcal{H}):(1-%
\mathrm{P})\mathrm{K}\mathrm{P}=0,\ \mathrm{P}\in \mathcal{P}\},
\label{eq:50}
\end{equation}%
\begin{equation*}
\mathcal{B}^{+}(\mathcal{H})=\{\mathrm{K}\in \mathcal{B}(\mathcal{H}):%
\mathrm{P}\mathrm{K}(1-\mathrm{P})=0,\ \mathrm{P}\in \mathcal{P}\}
\end{equation*}%
and call an operator $\mathrm{K}\in \mathcal{B}^{+},\ \ (\mathrm{K}\in 
\mathcal{B}^{-})$ \textbf{up-triangle} (\textbf{down-triangle}) with respect
to the projector chain $\mathcal{P}.$ \ Denote also by $\mathcal{B}_{p}(%
\mathcal{H}),\ \ p\in \lbrack 1,\infty ],$ the so called Neumann-Shattin
ideals and put 
\begin{equation}
\mathcal{B}_{\infty }^{+}(\mathcal{H}):=\mathcal{B}_{\infty }(\mathcal{H}%
)\cap \mathcal{B}^{+}(\mathcal{H}),\ \ \mathcal{B}_{\infty }^{-}(\mathcal{H}%
):=\mathcal{B}_{\infty }(\mathcal{H})\cap \mathcal{B}^{-}(\mathcal{H}).
\label{eq:51}
\end{equation}%
Subject to Definition 3.4 Banach subspaces (\ref{eq:51}) are Volterrian,,
being closed in $\mathcal{B}_{\infty }(\mathcal{H})$ and satisfying the
condition 
\begin{equation}
\mathcal{B}_{\infty }^{+}(\mathcal{H})\cap \mathcal{B}_{\infty }^{-}(%
\mathcal{H})=\varnothing .  \label{eq:52}
\end{equation}%
Denote also by $\mathcal{P}^{+}\ (\mathcal{P}^{-})$ the corresponding
projectors of the linear space 
\begin{equation*}
\widetilde{\mathcal{B}}_{\infty }(\mathcal{H}):=\mathcal{B}_{\infty }^{+}(%
\mathcal{H})\oplus \mathcal{B}_{\infty }^{-}(\mathcal{H})\subset \mathcal{B}%
_{\infty }(\mathcal{H})
\end{equation*}%
upon $\mathcal{B}_{\infty }^{+}(\mathcal{H})(\ \mathcal{B}_{\infty }^{-}(%
\mathcal{H})),$ and call them after \cite{GK} by \textbf{transformators} of
a \textbf{triangle shear}. The transformators $\mathcal{P}^{+}$ and $%
\mathcal{P}^{-}$ are known \cite{GK} to be continuous operators in ideals $%
\mathcal{B}_{p}(\mathcal{H}),\ \ p\in \lbrack 1,\infty ].$ From definitions
above one gets that 
\begin{equation}
\mathcal{P}^{+}(\Phi )+\mathcal{P}^{-}(\Phi )=\Phi ,\ \ \mathcal{P}^{\pm
}(\Phi )=\tau \mathcal{P}^{\mp }\tau (\Phi )  \label{eq:53}
\end{equation}%
for any $\Phi \in \mathcal{B}(\mathcal{H}),$ where $\tau :\mathcal{B}_{p}(%
\mathcal{H})\ \rightarrow \ \mathcal{B}_{p}(\mathcal{H})$ is the standard
involution in $\mathcal{B}_{p}(\mathcal{H})$ acting as $\tau (\Phi ):=\Phi
^{\ast }.$

\begin{remark}
It is clear and important that transformators $\mathcal{P}^{+}$ \ and $%
\mathcal{P}^{-}$ \ strongly depend on a fixed projector chain $\mathcal{P}.$
\end{remark}

Put now, by definition, 
\begin{equation}
\mathcal{V}_{f}^{\pm }:=\ \{1+\mathrm{K}_{\pm }:\mathrm{K}_{\pm }\in 
\mathcal{B}_{\infty }^{\pm }(\mathcal{H})\}  \label{eq:54}
\end{equation}%
and 
\begin{equation}
\mathcal{V}_{f}:=\ \{\mathbf{\Omega }_{+}^{-1}\cdot \mathbf{\Omega }_{-}:%
\mathbf{\Omega }_{\pm }\in \mathcal{V}_{f}^{\pm }\}.  \label{eq:55}
\end{equation}%
It is easy to check that $\mathcal{V}_{f}^{+}$ and $\mathcal{V}_{f}^{-}$ are
subgroups of invertible operators from $Aut(\mathcal{H})\cap \mathcal{B}(%
\mathcal{H})$ and, moreover, $\mathcal{V}_{f}^{+}\cap \mathcal{V}_{f}^{-}=\{%
\mathrm{1}\}$. Consider also the following two operator sets: 
\begin{equation*}
\mathcal{W}:=\{\Phi \in \mathcal{B}_{\infty }(\mathcal{H}):\ \text{Ker}(%
\mathrm{1}+\mathrm{P}\Phi \mathrm{P})=\{0\},\text{ \ }\mathrm{P}\in \mathcal{%
P}\},
\end{equation*}%
\begin{equation}
\mathcal{W}_{f}:=\{\Phi \in \mathcal{B}_{\infty }(\mathcal{H}):\text{ \ }%
\mathbf{\Omega }:=\mathrm{1}+\Phi \in \mathcal{V}_{f}\},  \label{eq:56}
\end{equation}%
which are characterized by the following (see \cite{GK,My}) theorem.

\begin{theorem}[I.C. Gokhberg and M.G. Krein]
The following conditions hold:

\begin{itemize}
\item[i)] $\mathcal{W}_{f}\subset \mathcal{W}$;

\item[ii)] $\mathcal{B}_{\omega }(\mathcal{H})\cap \mathcal{W}\subset 
\mathcal{W}_{f}$ where $\mathcal{B}_{\omega }(\mathcal{H})\subset \mathcal{B}%
(\mathcal{H})$ is the so called Macaev ideal;

\item[iii)] for any $\Phi \in \mathcal{W}_{f}$ \ it is necessary and
sufficient that at least one of integrals 
\begin{equation*}
\mathcal{K}_{+}(\mathbf{\Omega })=-\int_{\mathcal{P}}\mathrm{d}\mathrm{P}%
\Phi \mathrm{P}(\mathrm{1}+\mathrm{P}\Phi \mathrm{P})^{-1},
\end{equation*}%
\begin{equation}
(\mathrm{1}+\mathrm{K}_{-}(\mathbf{\Omega }))^{-1}-\mathrm{1}=-\int_{%
\mathcal{P}}(\mathrm{1}+\mathrm{P}\Phi \mathrm{P})^{-1}\mathrm{P}\Phi 
\mathrm{d}\mathrm{P}  \label{eq:57}
\end{equation}%
is convergent in the uniform operator topology, and, moreover, if the one
integral of (\ref{eq:40}) is convergent then the another one is convergent
too;

\item[iiii)] the factorization representation 
\begin{equation}
\mathbf{\Omega }=\mathrm{1}+\Phi =(\mathrm{1}+\mathrm{K}_{+}(\mathbf{\Omega }%
))^{-1}(\mathrm{1}+\mathrm{K}_{-}(\mathbf{\Omega }))  \label{eq:58}
\end{equation}%
for $\ \Phi \in \mathcal{W}_{f}$ is satisfied.
\end{itemize}
\end{theorem}

The theorem above is still abstract since it doesn't take into account the
crucial relationship (\ref{eq:38}) relating the operators representation (%
\ref{eq:58}) with a given differential operator $L\in \mathcal{L}(\mathcal{H}%
)$. Thus, it is necessary to satisfy the condition (\ref{eq:38}). If this
condition is due to (\ref{eq:20}) and (\ref{eq:34}) satisfied, the following
crucial equalities%
\begin{equation}
(\mathrm{1}+\mathrm{K}_{+}(\mathbf{\Omega }))L(\mathrm{1}+\mathrm{K}_{+}(%
\mathbf{\Omega }))^{-1}=\tilde{L}=(\mathrm{1}+\mathrm{K}_{-}(\mathbf{\Omega }%
))L(\mathrm{1}+\mathrm{K}_{-}(\mathbf{\Omega }))^{-1}  \label{eq:59}
\end{equation}%
in $\mathcal{H}$ and the corresponding congruence relationships 
\begin{equation}
(\tilde{L}_{ext}\otimes \mathrm{1})\widehat{\mathrm{K}}_{\pm }=(\mathrm{1}%
\otimes L_{ext}^{\ast })\widehat{\mathrm{K}}_{\pm }  \label{eq:60}
\end{equation}%
in $\mathcal{H}_{+}\otimes \mathcal{H}_{+}$ hold. Here by $\mathrm{\hat{K}}%
_{\pm }\in \mathcal{H}_{-}\otimes \mathcal{H}_{-}$ we denoted the
corresponding kernels of Volterra operators $\mathrm{K}_{\pm }(\mathbf{%
\Omega })\in \mathcal{B}_{\infty }^{\pm }(\mathcal{H}).$ Since the
factorization (\ref{eq:58}) is unique, the corresponding kernels must a
priori satisfy the conditions (\ref{eq:59}) and (\ref{eq:60}). Thereby the
self-similar congruence condition must be solved with respect to a kernel $%
\hat{\Phi}\in \mathcal{H}_{-}\otimes \mathcal{H}_{-}$ corresponding to the
integral operator $\Phi \in \mathcal{B}_{\infty }(\mathcal{H}),$ and next,
must be found the corresponding unique factorization (\ref{eq:58}),
satisfying a priori condition (\ref{eq:59}) and (\ref{eq:60}).

3.7 To realize this scheme define preliminarily a unique positive Borel
finite measure on the Borel subsets $\Delta \subset \mathrm{Q}$ of the open
set $\mathrm{Q}\subset \mathbb{R}^{m},$ satisfying for any projector $%
\mathrm{P}_{x}\in \mathcal{P}_{x}$ of a chain $\mathcal{P}_{x},$ marked by a
running point $x\in \mathrm{Q,}$ the following condition 
\begin{equation}
(u,\mathrm{P}_{x}(\Delta )v)_{\mathcal{H}}=\int_{\Delta \subset Q}(u,%
\mathcal{X}_{x}(y)v)\mathrm{d}\mu _{\mathcal{P}_{x}}(y)  \label{eq:61}
\end{equation}%
for all $u,v\in \mathcal{H}_{+},$ where $\mathcal{X}_{x}:\mathrm{Q}\
\rightarrow \ \mathcal{B}_{2}(\mathcal{H}_{+},\mathcal{H}_{-})$ is for any $%
x\in \mathrm{Q}$ a measurable with respect to some Borel measure $\mu _{%
\mathcal{P}_{x}}$ on Borel subsets of $\mathrm{Q}$ operator-valued mapping
of Hilbert-Schmidt type. The representation (\ref{eq:61}) follows due the
reasoning similar to that in \cite{Be}, based on the standard Radon-Nikodym
theorem \cite{Be,DS}. This means in particular, that in the weak sense 
\begin{equation}
\mathrm{P}_{x}(\Delta )=\int_{\Delta }\mathcal{X}_{x}(y)\mathrm{d}\mu _{%
\mathcal{P}_{x}}(y)  \label{eq:62}
\end{equation}%
for any Borel set $\Delta \in \mathrm{Q}$ and a running point $x\in \mathrm{Q%
}$. Making use now of the weak representation (\ref{eq:62}) the integral
expression like $\mathrm{I}_{f,g}(x)=\int_{\mathcal{P}_{x}}f(\mathrm{P}_{x})%
\mathrm{d}\mathrm{P}_{x}g(\mathrm{P}_{x}),$ $x\in \mathrm{Q,}$ for any
continuous mappings $f,g:\mathcal{P}_{x}\ \rightarrow \ \mathcal{B}(\mathcal{%
H})$ can be, obviously, represented as 
\begin{equation}
\mathrm{I}_{f,g}(x)=\int_{\mathrm{Q}}f(\mathrm{P}(y))\mathcal{\chi }_{x}(y)g(%
\mathrm{P}(y))\mathrm{d}\mu _{\mathcal{P}_{x}}(y).  \label{eq:63}
\end{equation}%
Thereby for the Volterrian operators (\ref{eq:57}) one can get the following
expressions: 
\begin{equation*}
\mathrm{K}_{+,x}(\mathbf{\Omega })=-\int_{\mathrm{Q}}(\mathrm{1}+\mathrm{P}%
_{x}(y)\Phi \mathrm{P}_{x}(y))^{-1}\mathrm{P}_{x}(y)\Phi \mathrm{d}\mu _{%
\mathcal{P}_{x,+}}(y),
\end{equation*}%
\begin{equation}
(\mathrm{1}+\mathrm{K}_{+,x}(\mathbf{\Omega }))^{-1}=\mathrm{1}-\int_{%
\mathrm{Q}}\mathrm{d}\mu _{\mathcal{P}_{x,+}}(y)\Phi \mathrm{P}_{x}(y)(%
\mathrm{1}+\mathrm{P}_{_{x}}(y)\Phi \mathrm{P}_{x}(y))^{-1}  \label{eq:64}
\end{equation}%
for some Borel measure $\mu _{\mathcal{P}_{x,+}}$ on \textrm{Q }and\textrm{\ 
}a given operator $\Phi \in \mathcal{B}_{\infty }(\mathcal{H}).$ The first
expression of (\ref{eq:64}) can be written down for the corresponding
kernels $\mathrm{\hat{K}}_{+,x}(y)\in $ $\mathcal{H}_{-}\otimes \mathcal{H}%
_{-}$ as follows 
\begin{equation}
\mathrm{\hat{K}}_{+,x}(y)=-\int_{\sigma (L)\cap \overline{\sigma }(L^{\ast
})}\mathrm{d}\rho _{\sigma ,+}(\lambda )\tilde{\psi}_{\lambda }(x)\otimes
\varphi _{\lambda }(y),  \label{eq:65}
\end{equation}%
where, due to the representation (\ref{eq:37}) and Theorem 2.2. we put for
any running points $x,y$ and $x^{\prime }$ $\in \mathrm{Q}$ \ the following
convolution of two kernels: 
\begin{equation}
((\mathrm{1}+\mathrm{P}_{x}(x^{\prime })\Phi \mathrm{P}_{x}(x^{\prime
}))^{-1})\ast (\psi _{\lambda }(x^{\prime })\otimes \varphi _{\lambda
}(y)):=\ \tilde{\psi}_{\lambda }(x)\otimes \varphi _{\lambda }(y).
\label{eq:66}
\end{equation}%
for $\lambda \in \sigma (\tilde{L})\cap \bar{\sigma}(L^{\ast })$ and some $%
\tilde{\psi}_{\lambda }\in \mathcal{H}_{-}$ $.$ Taking now into account the
representation (\ref{eq:26}) at $s="+",$ from (\ref{eq:65}) one gets easily
that the elementary congruent kernel 
\begin{equation}
\widehat{\widetilde{\mathrm{Z}}}_{\lambda }=\tilde{\psi}_{\lambda }\otimes
\varphi _{\lambda }  \label{eq:67}
\end{equation}%
satisfies the important conditions $(\tilde{L}_{ext}\otimes \mathrm{1})%
\widehat{\widetilde{\mathrm{Z}}}_{\lambda }=\lambda \widehat{\widetilde{%
\mathrm{Z}}}_{\lambda }$ and $(\mathrm{1}\otimes L^{\ast })\widehat{%
\widetilde{\mathrm{Z}}}_{\lambda }=\lambda \widehat{\widetilde{\mathrm{Z}}}%
_{\lambda }$ for any $\lambda \in $ $\sigma (\tilde{L})$ $\cap \bar{\sigma}%
(L^{\ast }).$ Now for the operator $\mathrm{K}_{+}(\mathbf{\Omega })\in 
\mathcal{B}_{\infty }^{+}(\mathcal{H})$ one finds the following integral
representation 
\begin{equation}
\mathrm{K}_{+}(\mathbf{\Omega })=-\int_{S_{+,x}^{(m)}}\mathrm{d}%
y\int_{\sigma (\widetilde{L})\cap \bar{\sigma}(L^{\ast })}\mathrm{d}\rho
_{\sigma ,+}(\lambda )\tilde{\psi}_{\lambda }(x)\bar{\varphi}_{\lambda }^{%
\mathrm{\intercal }}(y)(\cdot ),  \label{eq:68}
\end{equation}%
satisfying, evidently the congruence condition (\ref{eq:20}), where we put,
by definition, 
\begin{equation}
\mathrm{d}\mu _{\mathcal{P}_{x,+}}(y)=\chi _{S_{+,x}^{(m)}}\mathrm{d}y,
\label{eq:69}
\end{equation}%
with $\chi _{S_{+,x}^{(m)}}$ being the characteristic function of the
support of the measure $\mathrm{d}\mu _{\mathcal{P}_{x}},$ that is $supp$ $\
\mu _{\mathcal{P}_{x,+}}:=S_{+,x}^{(m)}\in \mathcal{K}(\mathrm{Q}).$
Completely similar reasonings can be applied for describing the structure of
the second factorizing operator $\mathrm{K}_{-}(\mathbf{\Omega })\in 
\mathcal{B}_{\infty }^{-}(\mathcal{H}):$ 
\begin{equation}
\mathrm{K}_{-}(\mathbf{\Omega })=-\int_{S_{-,x}^{(m)}}\mathrm{d}%
y\int_{\sigma (\widetilde{L})\cap \bar{\sigma}(L^{\ast })}\mathrm{d}\rho
_{\sigma ,-}(\lambda )\tilde{\psi}_{\lambda }(x)\bar{\varphi}_{\lambda }^{%
\mathrm{\intercal }}(y)(\cdot ),  \label{eq:70}
\end{equation}%
where, by definition, $S_{-,x}^{(m)}\subset \mathrm{Q},$ $x\in \mathrm{Q,}$
is, as before, the support $supp\mu _{\mathcal{P}_{x,-}}:=S_{-,x}^{(m)}\in 
\mathcal{K}(\mathrm{Q})$ of the corresponding \ to the operator (\ref{eq:70}%
) finite Borel measure $\mu _{\mathcal{P}_{x,-}}$ defined on the Borel
subsets of \textrm{Q}$\subset \mathbb{R}^{m}.$

It is naturally to put now $x\in \partial S_{+,x}^{(m)}\cap \partial
S_{-,x}^{(m)},$ being an intrinsic point of the boundary $\partial
S_{+,x}^{(m)}\backslash \partial Q=-\partial S_{-,x}^{(m)}\backslash
\partial Q:=\sigma _{x}^{(m-1)}\in \mathcal{K}(\mathrm{Q}),$ where $\mathcal{%
K}(\mathrm{Q})$ is, as before, some singular simplicial complex generated by
the open set $\mathrm{Q}\subset \mathbb{R}^{m}.$ Thus, for our Fredholm
operator $\ \mathbf{\Omega }:=\mathrm{1}+\Phi \in \mathrm{V}_{f}$ \ the
corresponding factorization is written down as 
\begin{equation}
\mathbf{\Omega }=(\mathrm{1}+\mathrm{K}_{+}(\mathbf{\Omega }))^{-1}(\mathrm{1%
}+\mathrm{K}_{-}(\mathbf{\Omega })):=\mathbf{\Omega }_{+}^{-1}\mathbf{\Omega 
}_{-},  \label{eq:71}
\end{equation}%
where integral operators $\mathrm{K}_{\pm }(\mathbf{\Omega })\in \mathcal{B}%
_{\infty }^{\pm }(\mathcal{H})$ are given by expression (\ref{eq:68}) and (%
\ref{eq:70}) parametrized by a running intrinsic point $x\in \mathrm{Q}$.

\section{The differential-geometric structure of a Lagrangian identity and
related Delsarte transmutation operators}

\setcounter{equation}{0}4.1. In Section 3 above we have studied in detail
the spectral structure of Delsarte transmutation Volterrian operators $%
\mathbf{\Omega }_{\pm }\in Aut(\mathcal{H})\cap \mathcal{B(H)}$ factorizing
some Fredholm operator $\mathbf{\Omega =\Omega }_{+}^{-1}\mathbf{\Omega }%
_{-} $ and stated their relationships with the approach suggested in \cite%
{GK,My}. In particular, we demonstrated the existence of some Borel measures 
$\mu _{\mathcal{P}_{x,\pm }}$ localized upon hypersurfaces $S_{\pm
,x}^{(m)}\in \mathcal{K}(\mathrm{Q})$ and related naturally with the
corresponding integral operators $\mathrm{K}_{\pm }(\mathbf{\Omega }),$
whose kernels $\mathrm{\hat{K}}_{\pm }(\mathbf{\Omega })\in \mathcal{H}%
_{-}\otimes \mathcal{H}_{-}$ are congruent to a pair of given differential
operators $(L,\tilde{L})\subset \mathcal{L(H)},$ satisfying the
relationships (\ref{eq:30}). In what will follow below we shall study some
differential-geometric properties of the Lagrange identity naturally
associated with two Delsarte related differential operators $L$ and $\tilde{L%
}$ in $\mathcal{H}$ and describe by means of some specially constructed
integral operator kernels the corresponding Delsarte transmutation operators
exactly in the same spectral form as it was studied in Section 3 above.

Let a multi-dimensional linear differential operator $\mathrm{L}:\mathcal{%
H\rightarrow H}$ of order $n(\mathrm{L})\in \mathbb{Z}_{+}$ be of the form 
\begin{equation}
\mathrm{L}\mathcal{(}x|\partial \mathcal{)}:=\sum_{|\alpha |=0}^{n(\mathrm{L}%
)}a_{\alpha }(x)\frac{\partial ^{|\alpha |}}{\partial x^{\alpha }},
\label{1.1}
\end{equation}%
and defined on a dense domain $D(\mathrm{L})\subset \mathcal{H},$ where, as
usually, $\alpha \in \mathbb{Z}_{+}^{m}$ is a multi-index, $x\in \mathbb{R}%
^{m},$ and for brevity one assumes that coefficients $a_{\alpha }\in 
\mathcal{S(}\mathbb{R}^{m};End\mathbb{C}^{N}),$ $\alpha \in \mathbb{Z}%
_{+}^{m}.$ Consider the following easily derivable generalized Lagrangian
identity for the differential expression (\ref{1.1}) :

\begin{equation}
<\mathrm{L}^{\ast }\varphi ,\psi >-<\varphi ,\mathrm{L}\psi
>=\sum_{i=1}^{m}(-1)^{i+1}\frac{\partial }{\partial x_{i}}Z_{i}[\varphi
,\psi ],  \label{1.2}
\end{equation}%
where $(\varphi ,\psi )\in \mathcal{H}^{\ast }\times \mathcal{H},$ mappings $%
Z_{i}:\mathcal{H}^{\ast }\times \mathcal{H}\rightarrow \mathbb{C},$ $i=%
\overline{1,m},$ are semilinear due to the construction and $\mathrm{L}%
^{\ast }:\mathcal{H}^{\ast }\rightarrow \mathcal{H}^{\ast }$ is the
corresponding formally conjugated to (\ref{1.1}) differential expression,
that is 
\begin{equation*}
\mathrm{L}^{\ast }(x|\partial ):=\sum_{|\alpha |=0}^{n(\mathcal{L}%
)}(-1)^{|\alpha |}\frac{\partial ^{|\alpha |}}{\partial x^{\alpha }}\cdot 
\bar{a}_{\alpha }^{\intercal }(x).
\end{equation*}%
Having multiplied the identity (\ref{1.2}) by the usual oriented Lebesgue
measure $dx=\wedge _{j=\overrightarrow{1,m}}dx_{j},$ we get that $\ \ $%
\begin{equation}
<\mathrm{L}^{\ast }\varphi ,\psi >dx-<\varphi ,\mathrm{L}\psi
>dx=dZ^{(m-1)}[\varphi ,\psi ]  \label{1.3}
\end{equation}%
\ for all $(\varphi ,\psi )\in \mathcal{H}^{\ast }\times \mathcal{H},$ where 
\begin{equation}
Z^{(m-1)}[\varphi ,\psi ]:=\sum_{i=1}^{m}dx_{1}\wedge dx_{2}\wedge ...\wedge
dx_{i-1}\wedge Z_{i}[\varphi ,\psi ]dx_{i+1}\wedge ...\wedge dx_{m}
\label{1.4}
\end{equation}%
is an $(m-1)-$differential form on $\mathbb{R}^{m}.$

4.2. Consider now all such pairs $(\varphi (\lambda ),\psi (\mu ))\in 
\mathcal{H}_{0}^{\ast }\times \mathcal{H}_{0}\subset \mathcal{H}_{-}\times 
\mathcal{H}_{-},$ $\lambda ,\mu \in \Sigma ,$ where as before 
\begin{equation}
\mathcal{H}_{+}\subset \mathcal{H}\subset \mathcal{H}_{-}  \label{1.4a}
\end{equation}%
is the usual Gelfand triple of Hilbert spaces \cite{Be,BS} related with our
Hilbert-Schmidt rigged Hilbert space $\mathcal{H},$ $\Sigma \in \mathbb{C}%
^{p},$ $p\in \mathbb{Z}_{+},$ is some fixed measurable space of parameters
endowed with a finite Borel measure $\rho ,$ that the differential form (\ref%
{1.4}) is exact, that is there exists a set of $(m-2)-$differential forms $%
\Omega ^{(m-2)}[\varphi (\lambda ),\psi (\mu )]$ $\in \Lambda ^{m-2}(\mathbb{%
R}^{m};\mathbb{C}),$ $\lambda ,\mu \in \Sigma ,$ on $\mathbb{R}^{m}$
satisfying the condition 
\begin{equation}
Z^{(m-1)}[\varphi (\lambda ),\psi (\mu )]=d\Omega ^{(m-2)}[\varphi (\lambda
),\psi (\mu )].  \label{1.5}
\end{equation}%
A way to realize this condition is to take some closed subspaces $\mathcal{H}%
_{0}^{\ast }$ and $\ \mathcal{H}_{0}\subset \mathcal{H}_{-}$ as solutions to
the corresponding linear differential equations under some boundary
conditions:%
\begin{eqnarray*}
\mathcal{H}_{0} &:&=\{\psi (\lambda )\in \mathcal{H}_{-}:\mathrm{L}\psi
(\lambda )=0,\text{ \ }\psi (\lambda )|_{x\in \Gamma }=0,\text{ }\lambda \in
\Sigma \},\text{ } \\
\mathcal{H}_{0}^{\ast } &:&=\{\varphi (\lambda )\in \mathcal{H}_{-}^{\ast }:%
\mathrm{L}^{\ast }\varphi (\lambda )=0,\text{ \ }\varphi (\lambda )|_{x\in
\Gamma }=0,\text{ }\lambda \in \Sigma \}.
\end{eqnarray*}%
The triple (\ref{1.4a}) allows naturally to determine properly a set of
generalized eigenfunctions for extended operators $\mathrm{L}$ $,\mathrm{L}%
^{\ast }$ : $\mathcal{H}_{-}\rightarrow \mathcal{H}_{-},$ if $\ \Gamma
\subset \mathbb{R}^{m}$ is taken as some (n-1)-dimensional piece-wise smooth
hypersurface embedded into the configuration space $\mathbb{R}^{m}.$ There
can exist, evidently, situations \cite{LS,Le,Fa} when boundary conditions
are not necessary.

Let now $S_{\pm }(\sigma _{x}^{(m-2)},\sigma _{x_{0}}^{(m-2)})\in C_{m-1}(M;%
\mathbb{C})$ denote some two non-intersecting $(m-1)$-dimensional piece-wise
smooth hypersurfaces from the singular simplicial chain group $C_{m-1}(M;%
\mathbb{C})\subset \mathcal{K}(M)$ of some topological compactification $M:=%
\mathbb{\bar{R}}^{m},$ such that their boundaries are the same, that is $%
\partial S_{\pm }(\sigma _{x}^{(m-2)},$ $\sigma _{x_{0}}^{(m-2)})=\sigma
_{x}^{(m-2)}-\sigma _{x_{0}}^{(m-2)}$ \ and, additionally, $\partial
(S_{+}(\sigma _{x}^{(m-2)},\sigma _{x_{0}}^{(m-2)})\cup S_{-}(\sigma
_{x}^{(m-2)},\sigma _{x_{0}}^{(m-2)}))=\oslash ,$ where $\sigma _{x}^{(m-2)}$
and $\sigma _{x_{0}}^{(m-2)}\in C_{m-2}(\mathbb{R}^{m};\mathbb{C})$ are some 
$(m-2)$-dimensional homological cycles from a suitable chain complex $%
\mathcal{K}(M)$ parametrized yet formally by means of two points $x,x_{0}\in
M$ and related in some way with the chosen above hypersurface $\Gamma $ \ $%
\subset M.$ Then from (\ref{1.5}) based on the general Stokes theorem \cite%
{Go,Te,Wa,DeR} one respectively gets easily that 
\begin{equation*}
\int_{S_{\pm }(\sigma _{x}^{(m-2)},\sigma
_{x_{0}}^{(m-2)})}Z^{(m-1)}[\varphi (\lambda ),\psi (\mu )]=\int_{\partial
S_{\pm }(\sigma _{x}^{(m-2)},\sigma _{x_{0}}^{(m-2)})}\Omega
^{(m-2)}[\varphi (\lambda ),\psi (\mu )]=
\end{equation*}%
\begin{eqnarray}
&&\int_{\sigma _{x}^{(m-2)}}\Omega ^{(m-2)}[\varphi (\lambda ),\psi (\mu
)]-\int_{\sigma _{x_{0}}^{(m-2)}}\Omega ^{(m-2)}[\varphi (\lambda ),\psi
(\mu )]  \label{1.6} \\
&:&=\Omega _{x}(\lambda ,\mu )-\Omega _{x_{0}}(\lambda ,\mu ),  \notag
\end{eqnarray}%
\begin{equation*}
\int_{S_{\pm }(\sigma _{x}^{(m-2)},\sigma _{x_{0}}^{(m-2)})}\overset{\_}{Z}%
^{(m-1),\intercal }[\varphi (\lambda ),\psi (\mu )]=\int_{\partial S_{\pm
}(\sigma _{x}^{(m-2)},\sigma _{x_{0}}^{(m-2)})}\bar{\Omega}^{(m-2),\intercal
}[\varphi (\lambda ),\psi (\mu )]=
\end{equation*}%
\begin{eqnarray*}
&&\int_{\sigma _{x}^{(m-2)}}\bar{\Omega}^{(m-2),\intercal }[\varphi (\lambda
),\psi (\mu )]-\int_{\sigma _{x_{0}}^{(m-2)}}\bar{\Omega}^{(m-2),\intercal
}[\varphi (\lambda ),\psi (\mu )] \\
&:&=\Omega _{x}^{\circledast }(\lambda ,\mu )-\Omega _{x_{0}}^{\circledast
}(\lambda ,\mu )
\end{eqnarray*}%
for the set of functions $(\varphi (\lambda ),\psi (\mu ))\in \mathcal{H}%
_{0}^{\ast }\times \mathcal{H}_{0},$ $\lambda ,\mu \in \Sigma ,$ with
operator kernels $\Omega _{x}(\lambda ,\mu ),$ $\Omega _{x}^{\circledast
}(\lambda ,\mu )\ $and $\ \Omega _{x_{0}}(\lambda ,\mu ),$ $\Omega
_{x}^{\circledast }(\lambda ,\mu ),$ $\lambda ,\mu \in \Sigma ,$ acting
naturally in the Hilbert space $L_{2}^{(\rho )}(\Sigma ;\mathbb{C}).$These
kernels are assumed further to be nondegenerate in $L_{2}^{(\rho )}(\Sigma ;%
\mathbb{C})$ and satisfying the homotopy conditions%
\begin{equation*}
\underset{x\rightarrow x_{0}}{lim}\Omega _{x}(\lambda ,\mu )\ =\ \Omega
_{x_{0}}(\lambda ,\mu ),\text{ \ }\underset{x\rightarrow x_{0}}{lim}\Omega
_{x}^{\circledast }(\lambda ,\mu )=\ \Omega _{x_{0}}^{\circledast }(\lambda
,\mu ).
\end{equation*}

4.3. Define now actions of the following two linear Delsarte permutations
operators $\mathbf{\Omega }_{\pm }:\mathcal{H}\rightarrow \mathcal{H}$ and $%
\mathbf{\Omega }_{\pm }^{\circledast }:\mathcal{H}^{\ast }\rightarrow 
\mathcal{H}^{\ast }$ still upon a fixed set of functions $(\varphi (\lambda
),\psi (\mu ))\in \mathcal{H}_{0}^{\ast }\times \mathcal{H}_{0},$ $\lambda
,\mu \in \Sigma :$%
\begin{equation*}
\tilde{\psi}(\lambda )=\mathbf{\Omega }_{\pm }(\psi (\lambda )):=\underset{%
\Sigma }{\int }d\rho (\eta )\underset{\Sigma }{\int }d\rho (\mu )\psi (\eta
)\Omega _{x}^{-1}(\eta ,\mu )\Omega _{x_{0}}(\mu ,\lambda ),
\end{equation*}%
\begin{equation}
\tilde{\varphi}(\lambda )=\mathbf{\Omega }_{\pm }^{\circledast }(\varphi
(\lambda )):=\underset{\Sigma }{\int }d\rho (\eta )\underset{\Sigma }{\int }%
d\rho (\mu )\varphi (\eta )\Omega _{x}^{\circledast ,-1}(\mu ,\eta )\Omega
_{x_{0}}^{\circledast }(\lambda ,\mu ).  \label{1.7}
\end{equation}%
Making use of the expressions (\ref{1.7}), based on arbitrariness of the
chosen \ set of functions $(\varphi (\lambda ),\psi (\mu ))\in \mathcal{H}%
_{0}^{\ast }\times \mathcal{H}_{0},$ $\lambda ,\mu \in \Sigma ,$ we can
easily retrieve the corresponding operator expressions for operators $%
\mathbf{\Omega }_{\pm }$ and $\mathbf{\Omega }_{\pm }^{\circledast }:%
\mathcal{H}_{-}\mathcal{\rightarrow H}_{-},$ forcing the kernels $\Omega
_{x_{0}}(\lambda ,\mu )$\ and $\Omega _{x_{0}}^{\circledast }(\lambda ,\mu
), $ $\lambda ,\mu \in \Sigma ,$ to variate:%
\begin{eqnarray*}
\tilde{\psi}(\lambda ) &=&\underset{\Sigma }{\int }d\rho (\eta )\underset{%
\Sigma }{\int }d\rho (\mu )\psi (\eta )\Omega _{x}(\eta ,\mu )\Omega
_{x}^{-1}(\mu ,\lambda ) \\
&&-\underset{\Sigma }{\int }d\rho (\eta )\underset{\Sigma }{\int }d\rho (\mu
)\psi (\eta )\Omega _{x}^{-1}(\eta ,\mu )]\times \\
&&\times \int_{S_{\pm }(\sigma _{x}^{(m-2)},\sigma
_{x_{0}}^{(m-2)})}Z^{(m-1)}[\varphi (\mu ),\psi (\lambda )])
\end{eqnarray*}%
\begin{eqnarray*}
&=&\psi (\lambda )-\int_{\Sigma }d\rho (\eta )\int_{\Sigma }d\rho (\mu )%
\underset{}{\int_{\Sigma }d\rho (\nu )\int_{\Sigma }}d\rho (\xi )\psi (\eta
)\Omega _{x}^{-1}(\eta ,\nu )\times \\
&&\times \Omega _{x_{0}}(\nu ,\xi )]\Omega _{x_{0}}^{-1}(\xi ,\mu
)\int_{S_{\pm }(\sigma _{x}^{(m-2)},\sigma
_{x_{0}}^{(m-2)})}Z^{(m-1)}[\varphi (\mu ),\psi (\lambda )]
\end{eqnarray*}%
\begin{equation*}
=\psi (\lambda )-\int_{\Sigma }d\rho (\eta )\int_{\Sigma }d\rho (\mu )\tilde{%
\psi}(\eta )\Omega _{x_{0}}^{-1}(\eta ,\mu )]\int_{S_{\pm }(\sigma
_{x}^{(m-2)},\sigma _{x_{0}}^{(m-2)})}Z^{(m-1)}[\varphi (\mu ),\psi (\lambda
)]
\end{equation*}%
\begin{equation*}
=(\mathbf{1}-\int_{\Sigma }d\rho (\eta )\int_{\Sigma }d\rho (\mu )\tilde{\psi%
}(\eta )\Omega _{x_{0}}^{-1}(\eta ,\mu )\times
\end{equation*}%
\begin{equation*}
\times \int_{S_{\pm }(\sigma _{x}^{(m-2)},\sigma
_{x_{0}}^{(m-2)})}Z^{(m-1)}[\varphi (\mu ),(\cdot )])\text{ }\psi (\lambda
):=\mathbf{\Omega }_{\pm }\cdot \psi (\lambda );
\end{equation*}%
\begin{eqnarray*}
\tilde{\varphi}(\lambda ) &=&\int_{\Sigma }d\rho (\eta )\int_{\Sigma }d\rho
(\mu )\varphi (\eta )\Omega _{x}^{\circledast ,-1}(\mu ,\eta )\Omega
_{x}^{\circledast }(\lambda ,\mu ) \\
&&-\int_{\Sigma }d\rho (\eta )\int_{\Sigma }d\rho (\mu )\varphi (\eta
)\Omega _{x}^{\circledast ,-1}(\mu ,\eta )\int_{S_{\pm }(\sigma
_{x}^{(m-2)},\sigma _{x_{0}}^{(m-2)})}\bar{Z}^{(m-1),\intercal }[\varphi
(\lambda ),\psi (\mu )]
\end{eqnarray*}%
\begin{eqnarray*}
&=&\varphi (\lambda )-\int_{\Sigma }d\rho (\eta )\int_{\Sigma }d\rho (\nu )%
\underset{}{\int_{\Sigma }d\rho (\xi )\int_{\Sigma }}d\rho (\mu )\varphi
(\eta )\Omega _{x}^{\circledast ,-1}(\xi ,\eta )\times \\
&&\times \Omega _{x_{_{0}}}^{\circledast }(\nu ,\xi )\Omega
_{x_{0}}^{\circledast ,-1}(\mu ,\nu )\int_{S_{\pm }(\sigma
_{x}^{(m-2)},\sigma _{x_{0}}^{(m-2)})}\bar{Z}^{(m-1),\intercal }[\varphi
(\lambda ),\psi (\mu )]
\end{eqnarray*}%
\begin{equation}
=(\mathbf{1}-\int_{\Sigma }d\rho (\eta )\int_{\Sigma }d\rho (\mu )\tilde{%
\varphi}(\eta )\Omega _{x_{_{0}}}^{\circledast ,-1}(\mu ,\eta )\times
\label{1.8}
\end{equation}%
\begin{equation*}
\times \int_{S_{\pm }(\sigma _{x}^{(m-2)},\sigma _{x_{0}}^{(m-2)})}\bar{Z}%
^{(m-1),\intercal }[(\cdot ),\psi (\mu )])\text{ }\varphi (\lambda ):=%
\mathbf{\Omega }_{\pm }^{\circledast }\cdot \varphi (\lambda ),
\end{equation*}%
where, by definition, 
\begin{equation*}
\mathbf{\Omega }_{\pm }:=\mathbf{1}-\int_{\Sigma }d\rho (\eta )\int_{\Sigma
}d\rho (\mu )\tilde{\psi}(\eta )\Omega _{x_{0}}^{-1}(\eta ,\mu )\int_{S_{\pm
}(\sigma _{x}^{(m-2)},\sigma _{x_{0}}^{(m-2)})}Z^{(m-1)}[\varphi (\mu
),(\cdot )]
\end{equation*}%
\begin{equation}
\mathbf{\Omega }_{\pm }^{\circledast }:=\mathbf{1}-\int_{\Sigma }d\rho (\eta
)\int_{\Sigma }d\rho (\mu )\tilde{\varphi}(\eta )\Omega
_{x_{_{0}}}^{\circledast ,-1}(\mu ,\eta )\int_{S_{\pm }(\sigma
_{x}^{(m-2)},\sigma _{x_{0}}^{(m-2)})}\bar{Z}^{(m-1),\intercal }[(\cdot
),\psi (\mu )]  \label{1.9}
\end{equation}%
are of Volterra type multidimensional integral operators. It is to be noted
here that now elements $(\varphi (\lambda ),\psi (\mu ))\in \mathcal{H}%
_{0}^{\ast }\times \mathcal{H}_{0}$ and $(\tilde{\varphi}(\lambda ),\tilde{%
\psi}(\mu ))\in \mathcal{\tilde{H}}_{0}^{\ast }\times \mathcal{\tilde{H}}%
_{0},$ $\lambda ,\mu \in \Sigma ,$ inside the operator expressions (\ref{1.9}%
) are not arbitrary but now fixed. Therefore, the operators (\ref{1.9})
realize an extension of their actions (\ref{1.7}) on a fixed pair of
functions $(\varphi (\lambda ),\psi (\mu ))\in \mathcal{H}_{0}^{\ast }\times 
\mathcal{H}_{0},$ $\lambda ,\mu \in \Sigma ,$ upon the whole functional
space $\mathcal{H}^{\ast }\times \mathcal{H}.$

4.4. Due to the symmetry of expressions (\ref{1.7}) and (\ref{1.9}) with
respect to two sets of functions $(\varphi (\lambda ),\psi (\mu ))\in 
\mathcal{H}_{0}^{\ast }\times \mathcal{H}_{0}$ and $(\tilde{\varphi}(\lambda
),\tilde{\psi}(\mu ))\in \mathcal{\tilde{H}}_{0}^{\ast }\times \mathcal{%
\tilde{H}}_{0},$ $\lambda ,\mu \in \Sigma ,$ it is very easy to state the
following lemma.

\begin{lemma}
Operators (\ref{1.9}) are bounded and invertible of Volterra type
expressions in $\mathcal{H}^{\ast }\times \mathcal{H}$ \ whose inverse are
given as follows:%
\begin{equation}
\mathbf{\Omega }_{\pm }^{-1}:=\mathbf{1}-\int_{\Sigma }d\rho (\eta
)\int_{\Sigma }d\rho (\mu )\psi (\eta )\tilde{\Omega}_{x_{0}}^{-1}(\eta ,\mu
)\int_{S_{\pm }(\sigma _{x}^{(m-2)},\sigma _{x_{0}}^{(m-2)})}Z^{(m-1)}[%
\tilde{\varphi}(\mu ),(\cdot )]  \label{1.10}
\end{equation}%
\begin{equation*}
\mathbf{\Omega }_{\pm }^{\circledast ,-1}:=\mathbf{1}-\int_{\Sigma }d\rho
(\eta )\int_{\Sigma }d\rho (\mu )\varphi (\eta )\Omega
_{x_{_{0}}}^{\circledast ,-1}(\mu ,\eta )\int_{S_{\pm }(\sigma
_{x}^{(m-2)},\sigma _{x_{0}}^{(m-2)})}\bar{Z}^{(m-1),\intercal }[(\cdot ),%
\tilde{\psi}(\mu )]
\end{equation*}%
where two sets of functions $(\varphi (\lambda ),\psi (\mu ))\in \mathcal{H}%
_{0}^{\ast }\times \mathcal{H}_{0}$ and $(\tilde{\varphi}(\lambda ),\tilde{%
\psi}(\mu ))\in \mathcal{\tilde{H}}_{0}^{\ast }\times \mathcal{\tilde{H}}%
_{0},$ $\lambda ,\mu \in \Sigma ,$ are taken arbitrary but fixed.
\end{lemma}

For the expressions (\ref{1.10}) to be compatible with mappings (\ref{1.7})
the following actions must hold:%
\begin{equation*}
\psi (\lambda )=\mathbf{\Omega }_{\pm }^{-1}\cdot \tilde{\psi}(\lambda
)=\int_{\Sigma }d\rho (\eta )\int_{\Sigma }d\rho (\mu )\tilde{\psi}(\eta )%
\tilde{\Omega}_{x}^{-1}(\eta ,\mu )]\tilde{\Omega}_{x_{0}}(\mu ,\lambda ),
\end{equation*}%
\begin{equation}
\varphi (\lambda )=\mathbf{\Omega }_{\pm }^{\circledast ,-1}\cdot \tilde{%
\varphi}(\lambda )=\int_{\Sigma }d\rho (\eta )\int_{\Sigma }d\rho (\mu )%
\tilde{\varphi}(\eta )\tilde{\Omega}_{x}^{\circledast ,-1}(\mu ,\eta )\tilde{%
\Omega}_{x_{0}}^{\circledast }(\lambda ,\mu ),  \label{1.11}
\end{equation}%
where for any two sets of functions $(\varphi (\lambda ),\psi (\mu ))\in 
\mathcal{H}_{0}^{\ast }\times \mathcal{H}_{0}$ and $(\tilde{\varphi}(\lambda
),\tilde{\psi}(\mu ))\in \mathcal{\tilde{H}}_{0}^{\ast }\times \mathcal{%
\tilde{H}}_{0},$ $\lambda ,\mu \in \Sigma ,$ the next relationship is
satisfied:%
\begin{equation*}
(<\mathrm{\tilde{L}}^{\ast }\tilde{\varphi}(\lambda ),\tilde{\psi}(\mu )>-<%
\tilde{\varphi}(\lambda ),\mathrm{\tilde{L}}\tilde{\psi}(\mu )>)dx=d(\tilde{Z%
}^{(m-1)}[\tilde{\varphi}(\lambda ),\tilde{\psi}(\mu )]),
\end{equation*}%
\begin{equation}
\tilde{Z}^{(m-1)}[\tilde{\varphi}(\lambda ),\tilde{\psi}(\mu )]=d\tilde{%
\Omega}^{(m-2)}[\tilde{\varphi}(\lambda ),\tilde{\psi}(\mu )]  \label{1.12}
\end{equation}%
when 
\begin{equation}
\mathrm{\tilde{L}}:=\mathbf{\Omega }_{\pm }\mathrm{L}\mathbf{\Omega }_{\pm
}^{-1},\text{ \ }\mathrm{\tilde{L}}^{\ast }:=\mathbf{\Omega }_{\pm
}^{\circledast }\mathrm{L}^{\ast }\mathbf{\Omega }_{\pm }^{\circledast ,-1},%
\text{ }  \label{1.12a}
\end{equation}%
Moreover, the expressions above for $\mathrm{\tilde{L}}:\mathcal{H}%
\rightarrow \mathcal{H}$ and $\mathrm{\tilde{L}}^{\ast }:\mathcal{H}^{\ast
}\rightarrow \mathcal{H}^{\ast }$ don't depend on the choice of the indexes
below of operators $\mathbf{\Omega }_{+}$ or $\mathbf{\Omega }_{-}$ and are
in the result differential. Since the last condition determines properly
Delsarte transmutation operators (\ref{1.10}), we need to state the
following theorem.

\begin{theorem}
The pair $(\mathrm{\tilde{L},}$ $\mathrm{\tilde{L}}^{\ast })$ of operator
expressions $\mathrm{\tilde{L}}:=\mathbf{\Omega }_{\pm }\mathrm{L}\mathbf{%
\Omega }_{\pm }^{-1}$ and $\mathrm{\tilde{L}}^{\ast }:=\mathbf{\Omega }_{\pm
}^{\circledast }\mathrm{L}^{\ast }\mathbf{\Omega }_{\pm }^{\circledast ,-1}$
acting in the space $\mathcal{H}\times \mathcal{H}^{\ast }$ \ is purely
differential \ for any suitably chosen hyper-surfaces $\ S_{\pm }(\sigma
_{x}^{(m-2)},\sigma _{x_{0}}^{(m-2)})$ $\in C_{m-1}(M;\mathbb{C})$ \ from
the homology group $C_{m-1}(M;\mathbb{C}).$
\end{theorem}

\begin{proof}
For proving the theorem it is necessary to show that the formal
pseudo-differential expressions corresponding to operators $\mathrm{\tilde{L}%
}$ and $\mathrm{\tilde{L}}^{\ast }$ contain no integral elements. Making use
of an idea devised in \cite{SP,Ni}, one can formulate such a lemma.
\end{proof}

\begin{lemma}
A pseudo-differential operator $\mathrm{L}:\mathcal{H}\rightarrow \mathcal{H}
$ is purely differential iff the following equality \ 
\begin{equation}
(h,(\mathrm{L}\frac{\partial ^{|\alpha |}}{\partial x^{\alpha }})_{+}f)=(h,%
\mathrm{L}_{+}\frac{\partial ^{|\alpha |}}{\partial x^{\alpha }}f)
\label{1.13}
\end{equation}%
holds for any $|\alpha |\in \mathbb{Z}_{+}$ and all $(h,f)\in \mathcal{H}%
^{\ast }\times \mathcal{H},$ that is the condition (\ref{1.13}) is
equivalent to the equality $\mathrm{L}_{+}=\mathrm{L},$ where, as usually,
the sign $"(...)_{+}"$ \ means the purely differential part of the
corresponding expression inside the bracket.
\end{lemma}

Based now on this Lemma and exact expressions of operators (\ref{1.9}),
similarly to calculations done in \cite{SP}, one shows right away that
operators $\mathrm{\tilde{L}}$ and $\mathrm{\tilde{L}}^{\ast },$ depending
respectively only both on the homological cycles $\sigma _{x}^{(m-2)},\sigma
_{x_{0}}^{(m-2)}\in C_{m-2}(M;\mathbb{C})$ from a simplicial chain complex $%
\mathcal{K}(M),$ marked by points $x,x_{0}\in \mathbb{R}^{m},$ and on two
sets of functions $(\varphi (\lambda ),\psi (\mu ))\in \mathcal{H}_{0}^{\ast
}\times \mathcal{H}_{0}$ and $(\tilde{\varphi}(\lambda ),\tilde{\psi}(\mu
))\in \mathcal{\tilde{H}}_{0}^{\ast }\times \mathcal{\tilde{H}}_{0},$ $%
\lambda ,\mu \in \Sigma ,$ are purely differential thereby finishing the
proof.$\blacktriangleright $

The differential-geometric construction suggested above can be nontrivially
generalized for the case of $m\in \mathbb{Z}_{+}$ commuting to each other
differential operators in a Hilbert space $\mathcal{H}$ \ giving rise to a
new look at theory of Delsarte transmutation operators based on
differential-geometric and topological de Rham-Hodge techniques. These
aspects will be discussed in detail in the next two Sections below.\bigskip

\section{The general differential-geometric and topological structure of
Delsarte transmutation operators: the de Rham-Hodge-Skrypnik theory}

\setcounter{equation}{0}5.1. Below we shall explain the corresponding
differential-geometric and topological nature of these spectral related
results obtained above and generalize them to a set $\mathcal{L}$\ \ of
commuting differential operators Delsarte related with another commuting set 
$\mathcal{\tilde{L}}$ of differential operators in $\mathcal{H}.$ These
results are \ deeply based on \ the de Rham-Hodge-Skrypnik theory \cite%
{DeR,DeR1,Te,Sk,Wa} of special differential complexes giving rise to
effective analytical expressions for the corresponding Delsarte
transmutation Volterra type operators in a given Hilbert space $\mathcal{H}.$
As a by-product one obtains the integral operator structure of Delsarte
transmutation operators for polynomial pencils of differential operators in $%
\mathcal{H}$ having many applications both in spectral theory of such \
multidimensional operator pencils \cite{PSP,GPSP,PSPS,DSa} and in soliton
theory \cite{Ni,FT,No,PM,Ma} of multidimensional integrable dynamical
systems on functional manifolds, being very important for diverse
applications in modern mathematical physics.

Let $M:=\mathbb{\bar{R}}^{m}$ denote as before a suitably compactified
metric space of dimension $m=dimM\in \mathbb{Z}_{+}$ (without boundary) and
define some finite set $\mathcal{L}$ \ of smooth commuting to each other
linear differential operators 
\begin{equation}
\mathrm{L}_{j}(x|\partial ):=\sum_{|\alpha |=0}^{n(L_{j})}a_{\alpha
}^{(j)}(x)\partial ^{|\alpha |}/\partial x^{\alpha }  \label{3.1}
\end{equation}%
with respect to $x\in M,$ having smooth \ enough coefficients $a_{\alpha
}^{(j)}\in \mathcal{S}(M;End\mathbb{C}^{N}),$ $|\alpha |=\overline{0,n(%
\mathrm{L}_{j})},$ $n(\mathrm{L}_{j})\in \mathbb{Z}_{+},$ $j=\overline{1,m},$
and acting in the Hilbert space $\mathcal{H}:=L_{2}(M;\mathbb{C}^{N}).$ It
is assumed also that domains $D(\mathrm{L}_{j}):=D(\mathcal{L})\subset 
\mathcal{H},$ $j=\overline{1,m},$ are dense in $\mathcal{H}.$

Consider now a generalized external anti-differentiation operator $d_{%
\mathcal{L}}$ :$\Lambda (M;\mathcal{H)\rightarrow }\Lambda (M;\mathcal{H)}$
acting in the Grassmann algebra $\Lambda (M;\mathcal{H)}$ as follows: for
any $\beta ^{(k)}\in \Lambda ^{k}(M;\mathcal{H)},$ $k=\overline{0,m},$ 
\begin{equation}
d_{\mathcal{L}}\beta ^{(k)}:=\sum_{j=1}^{m}dx_{j}\wedge \mathrm{L}%
_{j}(x|\partial )\beta ^{(k)}\in \Lambda ^{k+1}(M;\mathcal{H)}.  \label{3.2}
\end{equation}%
It is easy to see that the operation (\ref{3.2}) in the case $\mathrm{L}%
_{j}(x|\partial ):=\partial /\partial x_{j},$ $j=\overline{1,m},$ coincides
exactly with the standard external differentiation $d=\sum_{j=1}^{m}dx_{j}%
\wedge \partial /\partial x_{j}$ on the Grassmann algebra $\Lambda (M;%
\mathcal{H)}.$ Making use of the operation (\ref{3.2}) on $\Lambda (M;%
\mathcal{H)},$ one can construct the following generalized de Rham co-chain
complex

\begin{equation}
\mathcal{H}\rightarrow \Lambda ^{0}(M;\mathcal{H)}\overset{d_{\mathcal{L}}}{%
\rightarrow }\Lambda ^{1}(M;\mathcal{H)}\overset{d_{\mathcal{L}}}{%
\rightarrow }...\overset{d_{\mathcal{L}}}{\rightarrow }\Lambda ^{m}(M;%
\mathcal{H)}\overset{d_{\mathcal{L}}}{\rightarrow }0.  \label{3.3}
\end{equation}%
The following important property concerning the complex (\ref{3.3}) holds.

\begin{lemma}
The co-chain complex (\ref{3.3}) is exact.
\end{lemma}

\begin{proof}
It follows easily from the equality $d_{\mathcal{L}}d_{\mathcal{L}}=0$
holding due to the commutation of operators (\ref{3.1}).$\triangleright $
\end{proof}

\bigskip

5.2. Below we will follow the ideas developed before in \cite{Sk,DeR1}. A
differential form $\beta \in \Lambda (M;\mathcal{H)}$ will be called $d_{%
\mathcal{L}}$-closed if $d_{\mathcal{L}}\beta =0,$ and a form $\gamma \in
\Lambda (M;\mathcal{H)}$ will be called $d_{\mathcal{L}}$-homological to
zero if there exists on $M$ such a form $\omega \in \Lambda (M;\mathcal{H)}$
that $\gamma =d_{\mathcal{L}}\omega .$

Consider now the standard algebraic Hodge star-operation 
\begin{equation}
\star :\Lambda ^{k}(M;\mathcal{H)\rightarrow }\Lambda ^{m-k}(M;\mathcal{H)},
\label{3.3a}
\end{equation}%
$k=\overline{0,m},$ as follows \cite{Ch,DeR,DeR1,Wa}: if $\beta \in \Lambda
^{k}(M;\mathcal{H)},$ then the form $\star \beta \in \Lambda ^{m-k}(M;%
\mathcal{H)}$ is such that:

i) $\ (m-k)$-dimensional volume $|\star \beta |$ of the form $\star \beta $
equals $k$-dimensional volume $|\beta |$ of the form $\beta ;$

ii) the $m$-dimensional measure $\bar{\beta}^{\intercal }\wedge \star \beta $
$>0$ under the fixed orientation on $M.$

Define also on the space $\Lambda (M;\mathcal{H)}$ the following natural
scalar product: for any $\beta ,\gamma \in \Lambda ^{k}(M;\mathcal{H)},$ $k=%
\overline{0,m},$%
\begin{equation}
(\beta ,\gamma ):=\int_{M}\bar{\beta}^{\intercal }\wedge \star \gamma .
\label{3.4}
\end{equation}%
Subject to the scalar product (\ref{3.4}) we can naturally construct the
corresponding Hilbert space 
\begin{equation*}
\mathcal{H}_{\Lambda }(M):=\overset{m}{\underset{k=0}{\oplus }}\mathcal{H}%
_{\Lambda }^{k}(M)
\end{equation*}%
well suitable for our further consideration. Notice also here that the Hodge
star $\star $-operation satisfies the following easily checkable property:
for any $\beta ,\gamma \in \mathcal{H}_{\Lambda }^{k}(M),$ $k=\overline{0,m}%
, $%
\begin{equation}
(\beta ,\gamma )=(\star \beta ,\star \gamma ),  \label{3.5}
\end{equation}%
that is the Hodge operation $\star :\mathcal{H}_{\Lambda }(M)\mathcal{%
\rightarrow H}_{\Lambda }(M)$ is isometry and its standard adjoint with
respect to the scalar product (\ref{3.4}) operation satisfies the condition $%
(\star )^{^{\prime }}=(\star )^{-1}.$

Denote by $d_{\mathcal{L}}^{\prime }$ the formally adjoint expression to the
external weak differential operation $d_{\mathcal{L}}$ :$\mathcal{H}%
_{\Lambda }(M)\mathcal{\rightarrow H}_{\Lambda }(M)$ in the Hilbert space $%
\mathcal{H}_{\Lambda }(M).$ Making now use of the operations $d_{\mathcal{L}%
}^{\prime }$ and $d_{\mathcal{L}}$ in $\mathcal{H}_{\Lambda }(M)$ one can
naturally define \cite{Ch,DeR1,Wa} the generalized Laplace-Hodge operator $%
\Delta _{\mathcal{L}}:\mathcal{H}_{\Lambda }(M)\rightarrow \mathcal{H}%
_{\Lambda }(M)$ as 
\begin{equation}
\Delta _{\mathcal{L}}:=d_{\mathcal{L}}^{\prime }d_{\mathcal{L}}+d_{\mathcal{L%
}}d_{\mathcal{L}}^{\prime }.  \label{3.6}
\end{equation}%
Take a form $\beta \in \mathcal{H}_{\Lambda }(M)$ satisfying the equality 
\begin{equation}
\Delta _{\mathcal{L}}\beta =0.  \label{3.7}
\end{equation}%
Such a form is called \textit{harmonic}. One can also verify that a harmonic
form $\beta \in \mathcal{H}_{\Lambda }(M)$ satisfies simultaneously the
following two adjoint conditions:%
\begin{equation}
d_{\mathcal{L}}^{\prime }\beta =0,\text{ \ \ }d_{\mathcal{L}}\beta =0,
\label{3.8}
\end{equation}%
easily stemming from (\ref{3.6}) and (\ref{3.8}).

\bigskip\ It is not hard to check that the following differential operation
in $\mathcal{H}_{\Lambda }(M)$%
\begin{equation}
d_{\mathcal{L}}^{\ast }:=\star d_{\mathcal{L}}^{\prime }(\star )^{-1}
\label{3.9}
\end{equation}%
defines also a usual \cite{Go,Te,DeR} external anti-differential operation
in $\mathcal{H}_{\Lambda }(M).$ The corresponding dual to (\ref{3.3})
co-chain complex 
\begin{equation}
\mathcal{H}\rightarrow \Lambda ^{0}(M;\mathcal{H)}\overset{d_{\mathcal{L}%
}^{\ast }}{\rightarrow }\Lambda ^{1}(M;\mathcal{H)}\overset{d_{\mathcal{L}%
}^{\ast }}{\rightarrow }...\overset{d_{\mathcal{L}}^{\ast }}{\rightarrow }%
\Lambda ^{m}(M;\mathcal{H)}\overset{d_{\mathcal{L}}^{\ast }}{\rightarrow }0
\label{3.9a}
\end{equation}%
is evidently exact too, as the property $d_{\mathcal{L}}^{\ast }d_{\mathcal{L%
}}^{\ast }=0$ holds due to the definition (\ref{3.9}).

\bigskip 5.3. Denote further by $\mathcal{H}_{\Lambda (\mathcal{L})}^{k}(M),$
$k=\overline{0,m},$ the co$\hom $o$\log $y groups of $d_{\mathcal{L}}$%
-closed and by $\mathcal{H}_{\Lambda (\mathcal{L}^{\ast })}^{k}(M),$ $k=%
\overline{0,m},$ the co$\hom $o$\log $y groups of $d_{\mathcal{L}}^{\ast }$%
-closed differential forms, respectively, and by $\mathcal{H}_{\Lambda (%
\mathcal{L}^{\ast }\mathcal{L})}^{k}(M),$ $k=\overline{0,m},$ the abelian
groups of harmonic differential forms from the Hilbert sub-spaces $\mathcal{H%
}_{\Lambda }^{k}(M),$ $k=\overline{0,m}.$ Before formulating next results,
define the standard Hilbert-Schmidt rigged chain \cite{Be} of positive and
negative Hilbert spaces \ of differential forms 
\begin{equation}
\mathcal{H}_{\Lambda ,+}^{k}(M)\subset \mathcal{H}_{\Lambda }^{k}(M)\subset 
\mathcal{H}_{\Lambda ,-}^{k}(M),  \label{3.9b}
\end{equation}%
the corresponding rigged chains of Hilbert sub-spaces for harmonic forms 
\begin{equation}
\mathcal{H}_{\Lambda (\mathcal{L}^{\ast }\mathcal{L}),+}^{k}(M)\subset 
\mathcal{H}_{\Lambda (\mathcal{L}^{\ast }\mathcal{L})}^{k}(M)\subset 
\mathcal{H}_{\Lambda (\mathcal{L}^{\ast }\mathcal{L}),-}^{k}(M)  \label{3.9c}
\end{equation}%
and cohomology groups:%
\begin{eqnarray}
\mathcal{H}_{\Lambda (\mathcal{L}),+}^{k}(M) &\subset &\mathcal{H}_{\Lambda (%
\mathcal{L})}^{k}(M)\subset \mathcal{H}_{\Lambda (\mathcal{L}),-}^{k}(M),
\label{3.9d} \\
\mathcal{H}_{\Lambda (\mathcal{L}^{\ast }),+}^{k}(M) &\subset &\mathcal{H}%
_{\Lambda (\mathcal{L}^{\ast })}^{k}(M)\subset \mathcal{H}_{\Lambda (%
\mathcal{L}^{\ast }),-}^{k}(M)  \notag
\end{eqnarray}%
for any $k=\overline{0,m}.$ Assume also that the Laplace-Hodge operator (\ref%
{3.6}) is elliptic in $\mathcal{H}_{\Lambda }^{0}(M).$ Now by reasonings
similar to those in \cite{Ch,Te,DeR,Wa} one can formulate the following a
little generalized de Rham-Hodge theorem.

\begin{theorem}
The groups of harmonic forms $\mathcal{H}_{\Lambda (\mathcal{L}^{\ast }%
\mathcal{L}),-}^{k}(M),$ $k=\overline{0,m},$ are, respectively, isomorphic
to the cohomology groups $(H^{k}(M;\mathbb{C}))^{|\Sigma |},$ $k=\overline{%
0,m},$ where $H^{k}(M;\mathbb{C)}$ is the $k-$th cohomology group of the
manifold $M$ with complex coefficients, $\Sigma \subset $ $\mathbb{C}^{p},$ $%
p\in \mathbb{Z}_{+},$ $|\Sigma |:=card$ $\Sigma ,$ is a set of suitable
"spectral" parameters marking the linear space of independent $d_{\mathcal{L}%
}^{\ast }$-closed 0-forms from $\mathcal{H}_{\Lambda (\mathcal{L}),-}^{0}(M)$
and, moreover, the following direct sum decompositions 
\begin{equation}
\mathcal{H}_{\Lambda (\mathcal{L}^{\ast }\mathcal{L}),-}^{k}(M)\oplus \Delta 
\mathcal{H}_{\Lambda ,-}^{k}(M)=\mathcal{H}_{\Lambda ,-}^{k}(M)=\mathcal{H}%
_{\Lambda (\mathcal{L}^{\ast }\mathcal{L}),-}^{k}(M)\oplus d_{\mathcal{L}}%
\mathcal{H}_{\Lambda ,-}^{k-1}(M)\oplus d_{\mathcal{L}}^{^{\prime }}\mathcal{%
H}_{\Lambda ,-}^{k+1}(M)  \label{3.9e}
\end{equation}%
hold for any $k=\overline{0,m}.$
\end{theorem}

Another variant of the statement similar to that above was formulated in 
\cite{Sk,Sk1,Sk2,Sk3} and reads as the following generalized de
Rham-Hodge-Skrypnik theorem.

\begin{theorem}
(See Skrypnik I.V. \cite{Sk}) \ The generalized cohomology groups $\mathcal{H%
}_{\Lambda (\mathcal{L}),-}^{k}(M),k=\overline{0,m},$ are isomorphic,
respectively, to the cohomology groups $(H^{k}(M;\mathbb{C}))^{|\Sigma |},$ $%
k=\overline{0,m}.$
\end{theorem}

\begin{proof}
A proof of this theorem is based on some special sequence \cite{Sk} of
differential Lagrange type identities.
\end{proof}

Define the following closed subspace 
\begin{equation}
\mathcal{H}_{0}^{\ast }:=\{\varphi ^{(0)}(\lambda )\in \mathcal{H}_{\Lambda (%
\mathcal{L}^{\ast }),-}^{0}(M):d_{\mathcal{L}}^{\ast }\varphi ^{(0)}(\lambda
)=0,\text{ }\varphi ^{(0)}(\lambda )|_{\Gamma }=0,\text{ }\lambda \in \Sigma
\}  \label{3.10}
\end{equation}%
for some smooth $(m-1)$-dimensional hypersurface $\Gamma \subset M$ and $%
\Sigma \subset (\sigma (\mathcal{L})\cap \bar{\sigma}(\mathcal{L}^{\ast
}))\times \Sigma _{\sigma }\subset \mathbb{C}^{p},$ where $\mathcal{H}%
_{\Lambda (\mathcal{L}^{\ast }),-}^{0}(M)$ is, as above, a suitable
Hilbert-Schmidt rigged \cite{Be,BS} zero-order cohomology group space from
the co-chain given by (\ref{3.9d}), $\sigma (\mathcal{L})$ and $\sigma (%
\mathcal{L}^{\ast })$ are, respectively, mutual spectra of the sets of
commuting operators $\mathcal{L}$ and $\mathcal{L}^{\ast }$ in $\mathcal{H}.$
Thereby the dimension $\dim $ $\mathcal{H}_{0}^{\ast }=card$ $\Sigma $ is
assumed to be known. The next lemma stated by Skrypnik I.V. being
fundamental for the proof holds.

\begin{lemma}
(See Skrypnik\ I.V. \cite{Sk,Sk1,Sk2,Sk3}) There exists a set of
differential $(k+1)$-forms $Z^{(k+1)}[\varphi ^{(0)}(\lambda ),d_{\mathcal{L}%
}\psi ^{(k)}]\in \Lambda ^{k+1}(M;\mathbb{C}),$ $k=\overline{0,m},$ and a
set of $k$-forms $Z^{(k)}[\varphi ^{(0)}(\lambda ),\psi ^{(k)}]\in \Lambda
^{k}(M;\mathbb{C}),$ $k=\overline{0,m},$ parametrized by a set $\Sigma \ni
\lambda $ and semilinear in $(\varphi ^{(0)}(\lambda ),\psi ^{(k)})\in 
\mathcal{H}_{0}^{\ast }\times \mathcal{H}_{\Lambda ,-}^{k}(M),$ such that%
\begin{equation}
Z^{(k+1)}[\varphi ^{(0)}(\lambda ),d_{\mathcal{L}}\psi
^{(k)}]=dZ^{(k)}[\varphi ^{(0)}(\lambda ),\psi ^{(k)}]  \label{3.11}
\end{equation}%
for all $k=\overline{0,m}$ \ and $\lambda \in \Sigma .$
\end{lemma}

\begin{proof}
A proof is based on the following generalized Lagrange type identity holding
for any pair $(\varphi ^{(0)}(\lambda ),\psi ^{(k)})\in \mathcal{H}%
_{0}^{\ast }\times \mathcal{H}_{\Lambda ,-}^{k}(M):$%
\begin{eqnarray}
0 &=&<d_{\mathcal{L}}^{\ast }\varphi ^{(0)}(\lambda ),\star (\psi
^{(k)}\wedge \bar{\gamma})>  \label{3.12} \\
&:&=<\star d_{\mathcal{L}}^{\prime }(\star )^{-1}\varphi ^{(0)}(\lambda
),\star (\psi ^{(k)}\wedge \bar{\gamma})>  \notag \\
&=&<d_{\mathcal{L}}^{\prime }(\star )^{-1}\varphi ^{(0)}(\lambda ),\psi
^{(k)}\wedge \bar{\gamma}>=<(\star )^{-1}\varphi ^{(0)}(\lambda ),d_{%
\mathcal{L}}\psi ^{(k)}\wedge \bar{\gamma}>  \notag
\end{eqnarray}%
\begin{eqnarray*}
&&+Z^{(k+1)}[\varphi ^{(0)}(\lambda ),d_{\mathcal{L}}\psi ^{(k)}]\wedge \bar{%
\gamma}\text{ } \\
&=&<(\star )^{-1}\varphi ^{(0)}(\lambda ),d_{\mathcal{L}}\psi ^{(k)}\wedge 
\bar{\gamma}>+dZ^{(k)}[\varphi ^{(0)}(\lambda ),\psi ^{(k)}]\wedge \bar{%
\gamma},
\end{eqnarray*}%
where $Z^{(k+1)}[\varphi ^{(0)}(\lambda ),d_{\mathcal{L}}\psi ^{(k)}]\in
\Lambda ^{k+1}(M;\mathbb{C}),$ $k=\overline{0,m},$ and $Z^{(k)}[\varphi
^{(0)}(\lambda ),\psi ^{(k)}]\in \Lambda ^{k-1}(M;\mathbb{C}),$ $k=\overline{%
0,m},$ are some semilinear differential forms parametrized by a parameter $%
\lambda \in \Sigma ,$ and $\bar{\gamma}\in \Lambda ^{m-k-1}(M;\mathbb{C})$
is arbitrary constant $(m-k-1)$-form. Thereby, the semilinear differential $%
(k+1)$-forms $Z^{(k+1)}[\varphi ^{(0)}(\lambda ),d_{\mathcal{L}}\psi
^{(k)}]\in \Lambda ^{k+1}(M;\mathbb{C}),$ $k=\overline{0,m},$ and $k$-forms $%
Z^{(k)}[\varphi ^{(0)}(\lambda ),\psi ^{(k)}]\in \Lambda ^{k}(M;\mathbb{C}),$
$k=\overline{0,m},$ $\lambda \in \Sigma ,$ constructed above exactly
constitute those searched for in the Lemma.$\triangleright $
\end{proof}

Based now on this Lemma 3.3 one can construct the cohomology group
isomorphism claimed in Theorem 3.2 formulated above. Namely, following \cite%
{Sk,Sk1,Sk3,Sk2,Lo}, let us take some singular simplicial \cite{Te,DeR,Wa,HK}
complex $\mathcal{K}(M)$ of the manifold $M$ and introduce linear mappings $%
B_{\lambda }^{(k)}:\mathcal{H}_{\Lambda ,-}^{k}(M)\rightarrow C_{k}(M;%
\mathbb{C})),$ $k=\overline{0,m},$ $\lambda \in \Sigma ,$ where $C_{k}(M;%
\mathbb{C}),$ $k=\overline{0,m},$ are as before free abelian groups over the
field $\mathbb{C}$ generated, respectively, by all $k$-chains of simplexes $%
S^{(k)}\in C_{k}(M;\mathbb{C}),$ $k=\overline{0,m},$ from the singular
simplicial complex $\mathcal{K}(M),$ as follows:%
\begin{equation}
B_{\lambda }^{(k)}(\psi ^{(k)}):=\sum_{S^{(k)}\in C_{k}(M;\mathbb{C}%
))}S^{(k)}\int_{S^{(k)}}Z^{(k)}[\varphi ^{(0)}(\lambda ),\psi ^{(k)}]
\label{3.13}
\end{equation}%
with $\psi ^{(k)}\in \mathcal{H}_{\Lambda ,-}^{k}(M),$ $k=\overline{0,m}.$
The following theorem based on mappings (\ref{3.13}) holds.

\begin{theorem}
(See Skrypnik I.V. \cite{Sk} ) The set of operations (\ref{3.13})
parametrized by $\lambda \in \Sigma $ realizes the cohomology groups
isomorphism formulated in Theorem 5.3.
\end{theorem}

\begin{proof}
A proof of this theorem one can get passing over in (\ref{3.13}) to the
corresponding cohomology $\mathcal{H}_{\Lambda (\mathcal{L}),-}^{k}(M)$ and
homology $H_{k}(M;\mathbb{C})$ groups of $M$ $\ $for every $k=\overline{0,m}%
. $ If one to take an element $\psi ^{(k)}:=\psi ^{(k)}(\mu )\in \mathcal{H}%
_{\Lambda (\mathcal{L}),-}^{k}(M),$ $k=\overline{0,m},$ solving the equation 
$d_{\mathcal{L}}\psi ^{(k)}(\mu )=0$ with $\mu \in \Sigma _{k}$ being some
set of \ the related "spectral" parameters marking elements of the subspace $%
\mathcal{H}_{\Lambda (\mathcal{L}),-}^{k}(M),$ then one finds easily from (%
\ref{3.13}) and the identity (\ref{3.12}) that 
\begin{equation}
dZ^{(k)}[\varphi ^{(0)}(\lambda ),\psi ^{(k)}(\mu )]=0  \label{3.14}
\end{equation}%
for all pairs $(\lambda ,\mu )\in \Sigma \times \Sigma _{k},$ $k=\overline{%
0,m}.$ This, in particular, means due to the Poincare lemma \cite%
{Go,Te,Wa,DeR} that there exist differential $(k-1)$-forms $\Omega
^{(k-1)}[\varphi ^{(0)}(\lambda ),\psi ^{(k)}(\mu )]\in \Lambda ^{k-1}(M;%
\mathbb{C}),$ $k=\overline{0,m},$ such that 
\begin{equation}
Z^{(k)}[\varphi ^{(0)}(\lambda ),\psi ^{(k)}(\mu )]=d\Omega ^{(k-1)}[\varphi
^{(0)}(\lambda ),\psi ^{(k)}(\mu )]  \label{3.15}
\end{equation}%
for all pairs $(\varphi ^{(0)}(\lambda ),\psi ^{(k)}(\mu ))\in \mathcal{H}%
_{0}^{\ast }\times \mathcal{H}_{\Lambda (\mathcal{L}),-}^{k}(M)$
parametrized by $\ (\lambda ,\mu )\in \Sigma \times \Sigma _{k},$ $k=%
\overline{0,m}.$ As a result of passing on the right-hand side of (\ref{3.13}%
) to the homology groups $H_{k}(M;\mathbb{C}),$ $k=\overline{0,m},$ one gets
due to the standard Stokes theorem \cite{Go,Wa,DeR,Te} that the mappings 
\begin{equation}
B_{\lambda }^{(k)}:\mathcal{H}_{\Lambda (\mathcal{L}),-}^{k}(M)%
\rightleftarrows H_{k}(M;\mathbb{C})  \label{3.16}
\end{equation}%
are isomorphisms for every $\lambda \in \Sigma $ and $\lambda \in \Sigma .$
Making further use of the Poincare duality \cite{Te,DeR,Wa,HK} between the
homology groups $H_{k}(M;\mathbb{C}),$ $k=\overline{0,m},$ and the
cohomology groups $H^{k}(M;\mathbb{C}),$ $k=\overline{0,m},$ respectively,
one obtains finally the statement claimed in theorem 3.5, that is $\mathcal{H%
}_{\Lambda (\mathcal{L}),-}^{k}(M)\simeq (H^{k}(M;\mathbb{C}))^{|\Sigma
|}.\triangleright $
\end{proof}

\bigskip

5.4. Assume now that $M:=\mathrm{T}^{r}\times \mathbb{\bar{R}}^{s},$ $%
dimM=s+r\in \mathbb{Z}_{+},$ and $\mathcal{H}:=L_{2}(\mathrm{T}^{r};L_{2}(%
\mathbb{R}^{s};\mathbb{C}^{N})),$ where $\mathrm{T}^{r}:=$ $\overset{r}{%
\underset{j=1}{\times }\mathrm{T}_{j}},$ $\mathrm{T}_{j}:=[0,T_{j})\subset 
\mathbb{R}_{+},$ $j=\overline{1,r},$ and put 
\begin{equation}
d_{\mathcal{L}}=\sum_{j=1}^{r}dt_{j}\wedge \mathrm{L}_{j}(t;x|\partial ),%
\text{ }\mathrm{L}_{j}(t;x|\partial ):=\partial /\partial
t_{j}-L_{j}(t;x|\partial ),  \label{3.16a}
\end{equation}%
with 
\begin{equation}
\text{ \ }L_{j}(t;x|\partial )=\sum_{|\alpha |=0}^{n(L_{j})}a_{\alpha
}^{(j)}(t;x)\partial ^{|\alpha |}/\partial x^{\alpha },  \label{3.16b}
\end{equation}%
$j=\overline{1,r},$ being differential operations parametrically dependent
on $t\in \mathrm{T}^{r}$ and defined on dense subspaces $D(\mathrm{L}_{j})=D(%
\mathcal{L})\subset $ $L_{2}(\mathbb{R}^{s};\mathbb{C}^{N}),$ $j=\overline{%
1,r}.$ It is assumed also that operators $\mathrm{L}_{j}:\mathcal{H}%
\rightarrow \mathcal{H},$ $\ \ j=\overline{1,r},$ are commuting to each
other.

Take now such a fixed pair $(\varphi ^{(0)}(\lambda ),\psi ^{(0)}(\mu
)dx)\in \mathcal{H}_{0}^{\ast }\times \mathcal{H}_{\Lambda (\mathcal{L}%
),-}^{s}(M),$ parametrized by elements $(\lambda ,\mu )\in \Sigma \times
\Sigma ,$ for which due to both Theorem 5.5 and the \ Stokes theorem \cite%
{Go,Te,Wa,DeR,HK} the following equality 
\begin{equation}
B_{\lambda }^{(s)}(\psi ^{(0)}(\mu )dx)=S_{(t;x)}^{(s)}\int_{\partial
S_{(t;x)}^{(s)}}\Omega ^{(s-1)}[\varphi ^{(0)}(\lambda ),\psi ^{(0)}(\mu )dx]
\label{3.17}
\end{equation}%
holds, where $S_{(t;x)}^{(s)}\in C_{s}(M;\mathbb{C})$ is some fixed element
parametrized by an arbitrarily chosen point $(t;x)\in M\cap S_{(t;x)}^{(s)}.$
Consider the next integral expressions 
\begin{eqnarray}
\Omega _{(t;x)}(\lambda ,\mu ) &:&=\int_{\partial S_{(t;x)}^{(s)}}\Omega
^{(s-1)}[\varphi ^{(0)}(\lambda ),\psi ^{(0)}(\mu )dx],\text{ }  \notag \\
\Omega _{(t_{0};x_{0})}(\lambda ,\mu ) &:&=\int_{\partial
S_{(t_{0};x_{0})}^{(s)}}\Omega ^{(s-1)}[\varphi ^{(0)}(\lambda ),\psi
^{(0)}(\mu )dx],  \label{3.18}
\end{eqnarray}%
with a point $(t_{0};x_{0})\in M\cap S_{(t_{0};x_{0})}^{(s)}$ being taken
fixed, $\lambda ,\mu \in \Sigma ,$ and interpret them as the corresponding
kernels \cite{Be} of the integral invertible operators of Hilbert-Schmidt
type $\ \Omega _{(t;x)},\Omega _{(t_{0};x_{0})}:L_{2}^{(\rho )}(\Sigma ;%
\mathbb{C})\rightarrow L_{2}^{(\rho )}(\Sigma ;\mathbb{C}),$ where $\rho $ \
is some finite Borel measure on the parameter set $\Sigma .$ It assumes also
above that the boundaries $\partial S_{(t;x)}^{(s)}$ $:=\sigma
_{(t;x)}^{(s-1)}$and $\partial S_{(t_{0};x_{0})}^{(s)}:=\sigma
_{(t_{0};x_{0})}^{(s-1)}$ are taken homological to each other as $%
(t;x)\rightarrow (t_{0};x_{0})\in M.$ Define now the expressions 
\begin{equation}
\mathbf{\Omega }_{\pm }:\psi ^{(0)}(\eta )\rightarrow \tilde{\psi}%
^{(0)}(\eta )  \label{3.19}
\end{equation}%
for $\psi ^{(0)}(\eta )dx\in \mathcal{H}_{\Lambda (\mathcal{L}),-}^{s}(M),$ $%
\eta \in \Sigma ,$ and some $\tilde{\psi}^{(0)}(\eta )dx\in \mathcal{H}%
_{\Lambda ,-}^{s}(M),$ where, by definition, for any $\eta \in \Sigma $ 
\begin{equation}
\tilde{\psi}^{(0)}(\eta ):=\psi ^{(0)}(\eta )\cdot \Omega
_{(t;x)}^{-1}\Omega _{(t_{0};x_{0})}  \label{3.20}
\end{equation}%
\begin{equation*}
=\int_{\Sigma }d\rho (\mu )\int_{\Sigma }d\rho (\xi )\psi ^{(0)}(\mu )\Omega
_{(t;x)}^{-1}(\mu ,\xi )\Omega _{(t_{0};x_{0})}(\xi ,\eta ),
\end{equation*}%
being motivated by the expression (\ref{3.17})$.$ Suppose now that the
elements (\ref{3.20}) are ones being related to some another Delsarte
transformed cohomology group $\mathcal{H}_{\Lambda (\mathcal{\tilde{L}}%
),-}^{s}(M)$ that is the following condition 
\begin{equation}
d_{\mathcal{\tilde{L}}}\tilde{\psi}^{(0)}(\eta )dx=0\text{ }  \label{3.21}
\end{equation}%
for $\tilde{\psi}^{(0)}(\eta )dx\in \mathcal{H}_{\Lambda (\mathcal{\tilde{L}}%
),-}^{s}(M),$ $\eta \in \Sigma ,$ and some new external anti-differentiation
operation in $\mathcal{H}_{\Lambda ,-}(M)$%
\begin{equation}
d_{\mathcal{\tilde{L}}}:=\sum_{j=1}^{m}dx_{j}\wedge \mathrm{\tilde{L}}%
_{j}(t;x|\partial ),\text{ }\mathrm{\tilde{L}}_{j}(t;x|\partial ):=\partial
/\partial t_{j}-\tilde{L}_{j}(t;x|\partial )  \label{3.22}
\end{equation}%
holds, where expressions 
\begin{equation}
\tilde{L}_{j}(t;x|\partial )=\sum_{|\alpha |=0}^{n(L_{j})}\tilde{a}_{\alpha
}^{(j)}(t;x)\partial ^{|\alpha |}/\partial x^{\alpha },  \label{3.22a}
\end{equation}%
$j=\overline{1,r},$ are differential operations in $L_{2}(\mathbb{R}^{s};%
\mathbb{C}^{N})$ parametrically dependent on $t\in \mathrm{T}^{r}.$

5.5. Put now that 
\begin{equation}
\mathrm{\tilde{L}}_{j}:=\mathbf{\Omega }_{\pm }\mathrm{L}_{j}\mathbf{\Omega }%
_{\pm }^{-1}  \label{3.23}
\end{equation}%
for each $\ j=\overline{1,r},$ where $\mathbf{\Omega }_{\pm }:\mathcal{%
H\rightarrow }\mathcal{H}$ are the corresponding Delsarte transmutation
operators related with some elements $S_{\pm }(\sigma
_{(x;t)}^{(s-1)},\sigma _{(x_{0};t_{0)}}^{(s-1)})$ $\in C_{s}(M;\mathbb{C})$
related naturally with\ homological to each other boundaries $\partial
S_{(x;t)}^{(s)}=$ $\sigma _{(x;t)}^{(s-1)}$\ \ and $\partial
S_{(x_{0};t_{0)}}^{(s)}=\sigma _{(x_{0};t_{0})}^{(s-1)}.$ Since all of
operators $\mathrm{L}_{j}:\mathcal{H\rightarrow }\mathcal{H},$ $j=\overline{%
1,r},$ were taken commuting, the same property also holds for the
transformed operators (\ref{3.23}), that is $[\mathrm{\tilde{L}}_{j},\mathrm{%
\tilde{L}}_{k}]=0,$ $k,j=\overline{0,m}.$ The latter is, evidently,
equivalent due to (\ref{3.23}) to the following general expression:%
\begin{equation}
d_{\mathcal{\tilde{L}}}=\mathbf{\Omega }_{\pm }d_{\mathcal{L}}\mathbf{\Omega 
}_{\pm }^{-1}.  \label{3.24}
\end{equation}%
For the condition (\ref{3.24}) and (\ref{3.21}) to be satisfied, let us
consider the corresponding to (\ref{3.17}) expressions 
\begin{equation}
\tilde{B}_{\lambda }^{(s)}(\tilde{\psi}^{(0)}(\eta )dx)=S_{(t;x)}^{(s)}%
\tilde{\Omega}_{(t;x)}(\lambda ,\eta ),  \label{3.25}
\end{equation}%
related with the corresponding external differentiation (\ref{3.24}), where $%
S_{(t;x)}^{(s)}\in C_{s}(M;\mathbb{C})$ and $(\lambda ,\eta )\in \Sigma
\times \Sigma .$ Assume further that there are also defined mappings 
\begin{equation}
\mathbf{\Omega }_{\pm }^{\circledast }:\varphi ^{(0)}(\lambda )\rightarrow 
\tilde{\varphi}^{(0)}(\lambda )  \label{3.26}
\end{equation}%
with $\mathbf{\Omega }_{\pm }^{\circledast }:\mathcal{H}^{\ast }\mathcal{%
\rightarrow H}^{\ast }$ being some operators associated (but not necessary
adjoint!) with the corresponding Delsarte transmutation operators $\mathbf{%
\Omega }_{\pm }:\mathcal{H\rightarrow H}$ \ and satisfying the standard
relationships $\mathrm{\tilde{L}}_{j}^{\ast }:=\mathbf{\Omega }_{\pm
}^{\circledast }\mathrm{L}_{j}^{\ast }\mathbf{\Omega }_{\pm }^{\circledast
,-1},$ $j=\overline{1,r}.$ The proper Delsarte type operators $\mathbf{%
\Omega }_{\pm }:\mathcal{H}_{\Lambda (\mathcal{L}),-}^{0}(M)\rightarrow 
\mathcal{H}_{\Lambda (\mathcal{\tilde{L}}),-}^{0}(M)$ \ are related with two
different realizations of the action (\ref{3.20}) under the necessary
conditions \ 
\begin{equation}
d_{\mathcal{\tilde{L}}}\tilde{\psi}^{(0)}(\eta )dx=0,\text{ \ \ }d_{\mathcal{%
\tilde{L}}}^{\ast }\tilde{\varphi}^{(0)}(\lambda )=0,\text{ \ \ }
\label{3.27}
\end{equation}%
needed to be satisfied \ and meaning, evidently, that the embeddings $\tilde{%
\varphi}^{(0)}(\lambda )\in \mathcal{H}_{\Lambda (\mathcal{\tilde{L}}^{\ast
}),-}^{0}(M),$ $\lambda \in \Sigma ,$ and $\tilde{\psi}^{(0)}(\eta )dx\in 
\mathcal{H}_{\Lambda (\mathcal{\tilde{L}}),-}^{s}(M),$ $\eta \in \Sigma ,$ \
are satisfied$.$\ Now we need to formulate a lemma being important for the
conditions (\ref{3.27}) to hold.

\begin{lemma}
The following invariance property 
\begin{equation}
\tilde{Z}^{(s)}=\Omega _{(t_{0};x_{0})}\Omega _{(t;x)}^{-1}Z^{(s)}\Omega
_{(t;x)}^{-1}\Omega _{(t_{0};x_{0})}  \label{3.28}
\end{equation}%
holds for any $(t;x)$ and $(t_{0};x_{0})\in M.$
\end{lemma}

As a result of (\ref{3.28}) and the symmetry invariance between cohomology
spaces $\mathcal{H}_{\Lambda (\mathcal{L}),-}^{0}(M)$ and $\mathcal{H}%
_{\Lambda (\mathcal{\tilde{L}}),-}^{0}(M)$ one obtains the following pairs
of related mappings: 
\begin{eqnarray}
\psi ^{(0)} &=&\tilde{\psi}^{(0)}\tilde{\Omega}_{(t;x)}^{-1}\tilde{\Omega}%
_{(t_{0};x_{0})},\text{ \ }\varphi ^{(0)}=\tilde{\varphi}^{(0)}\tilde{\Omega}%
_{(t;x)}^{\circledast ,-1}\tilde{\Omega}_{(t_{0};x_{0})}^{\circledast },
\label{3.29} \\
\tilde{\psi}^{(0)} &=&\psi ^{(0)}\Omega _{(t;x)}^{-1}\Omega _{(t_{0};x_{0})},%
\text{ \ \ }\tilde{\varphi}^{(0)}=\varphi ^{(0)}\Omega _{(t;x)}^{\circledast
,-1}\Omega _{(t_{0};x_{0})}^{\circledast },  \notag
\end{eqnarray}%
where the integral operator kernels from $L_{2}^{(\rho )}(\Sigma ;\mathbb{C}%
)\otimes L_{2}^{(\rho )}(\Sigma ;\mathbb{C})$ are defined as 
\begin{eqnarray}
\tilde{\Omega}_{(t;x)}(\lambda ,\mu ) &:&=\int_{\sigma _{(t;x)}^{(s)}}\tilde{%
\Omega}^{(s-2)}[\tilde{\varphi}^{(0)}(\lambda ),\tilde{\psi}^{(0)}(\mu )dx],%
\text{ }  \label{3.29a} \\
\tilde{\Omega}_{(t;x)}^{\circledast }(\lambda ,\mu ) &:&=\int_{\sigma
_{(t;x)}^{(s)}}\overset{\_}{\tilde{\Omega}}^{(s-2),\intercal }[\tilde{\varphi%
}^{(0)}(\lambda ),\tilde{\psi}^{(0)}(\mu )dx]  \notag
\end{eqnarray}%
for all $(\lambda ,\mu )\in \Sigma \times \Sigma ,$ giving rise to finding
proper Delsarte transmutation operators ensuring the pure differential
nature of\ the transformed expressions (\ref{3.23}).

Note here also that due to (\ref{3.28}) and (\ref{3.29}) the following
operator property 
\begin{equation}
\Omega _{(t_{0;}x_{0})}\Omega _{(t;x)}^{-1}\Omega _{(t_{0;}x_{0})}+\tilde{%
\Omega}_{(t_{0;}x_{0})}\Omega _{(t;x)}^{-1}\Omega _{(t_{0;}x_{0})}=0
\label{3.30}
\end{equation}%
holds $\ $for any $\ (t_{0;}x_{0})$ and $(t;x)\in M$ meaning that $\tilde{%
\Omega}_{(t_{0;}x_{0})}=-\Omega _{(t_{0;}x_{0})}.$

5.6. One can now define similar to (\ref{3.10}) the additional closed and
dense in $\mathcal{H}_{\Lambda ,-}^{0}(M)$ three subspaces%
\begin{equation}
\mathcal{H}_{0}:=\{\psi ^{(0)}(\mu )\in \mathcal{H}_{\Lambda (\mathcal{L}%
),-}^{0}(M):d_{\mathcal{L}}\psi ^{(0)}(\mu )=0,\text{ \ \ }\psi ^{(0)}(\mu
)|_{\Gamma }=0,\text{ }\mu \in \Sigma \},  \notag
\end{equation}%
\begin{equation}
\mathcal{\tilde{H}}_{0}:=\{\tilde{\psi}^{(0)}(\mu )\in \mathcal{H}_{\Lambda (%
\widetilde{\mathcal{L}}),-}^{0}(M):d_{\widetilde{\mathcal{L}}}\tilde{\psi}%
^{(0)}(\mu )=0,\text{ \ \ }\tilde{\psi}^{(0)}(\mu )|_{\tilde{\Gamma}}=0,%
\text{ }\mu \in \Sigma \},  \label{3.31}
\end{equation}%
\begin{equation*}
\mathcal{\tilde{H}}_{0}^{\ast }:=\{\tilde{\varphi}^{(0)}(\eta )\in \mathcal{H%
}_{\Lambda (\mathcal{\tilde{L}}^{\ast }),-}^{0}(M):d_{\widetilde{\mathcal{L}}%
}^{\ast }\tilde{\psi}^{(0)}(\eta )=0,\text{ \ \ }\tilde{\varphi}^{(0)}(\eta
)|_{\tilde{\Gamma}}=0,\text{ }\eta \in \Sigma \},
\end{equation*}%
where $\Gamma $ and $\tilde{\Gamma}\subset M$ are some smooth $(s-1)$%
-dimensional hypersurfaces, and construct the actions 
\begin{equation}
\mathbf{\Omega }_{\pm }:\psi ^{(0)}\rightarrow \tilde{\psi}^{(0)}:=\psi
^{(0)}\Omega _{(t;x)}^{-1}\Omega _{(t;x)},\text{ \ \ \ }\mathbf{\Omega }%
_{\pm }^{\circledast }:\varphi ^{(0)}\rightarrow \tilde{\varphi}%
^{(0)}:=\varphi ^{(0)}\Omega _{(t;x)}^{\circledast ,-1}\Omega
_{(t_{0};x_{0})}^{\circledast }  \label{3.32}
\end{equation}%
on arbitrary but fixed pairs of elements $(\varphi ^{(0)}(\lambda ),\psi
^{(0}(\mu ))\in \mathcal{H}_{0}^{\ast }\times \mathcal{H}_{0},$ parametrized
by the set $\Sigma ,$ where by definition, one needs that all obtained pairs 
$(\tilde{\varphi}^{(0)}(\lambda ),\tilde{\psi}^{(0)}(\mu )dx),$ $\lambda
,\mu \in \Sigma ,$ belong to $\mathcal{H}_{\Lambda (\mathcal{\tilde{L}}%
^{\ast }),-}^{0}(M)\times \mathcal{H}_{\Lambda (\mathcal{\tilde{L}}%
),-}^{s}(M).$ Note also that related operator property (\ref{3.30}) can be
compactly written down as follows:%
\begin{equation}
\tilde{\Omega}_{(t;x)}=\tilde{\Omega}_{(t_{0};x_{0})}\Omega
_{(t;x)}^{-1}\Omega _{(t_{0};x_{0})}=-\Omega _{(t_{0};x_{0})}\Omega
_{(t;x)}^{-1}\Omega _{(t_{0};x_{0})}.  \label{3.33}
\end{equation}%
Construct now from the expressions (\ref{3.32}) the following operator
kernels from the Hilbert space $L_{2}^{(\rho )}(\Sigma ;\mathbb{C})\otimes
L_{2}^{(\rho )}(\Sigma ;\mathbb{C}):$%
\begin{eqnarray*}
&&\Omega _{(t;x)}(\lambda ,\mu )-\Omega _{(t_{0};x_{0})}(\lambda ,\mu
)=\int_{\partial S_{(t;x)}^{(s)}}\Omega ^{(s-1)}[\varphi ^{(0)}(\lambda
),\psi ^{(0)}(\mu )dx] \\
&:&-\int_{\partial S_{(t_{0};x_{0})}^{(s)}}\Omega ^{(s-1)}[\varphi
^{(0)}(\lambda ),\psi ^{(0)}(\mu )dx]
\end{eqnarray*}%
\begin{eqnarray}
&=&\underset{S_{\pm }^{(s)}(\sigma _{(t;x)}^{(s-1)},\sigma
_{(t_{0};x_{0})}^{(s-1)})}{\int }d\Omega ^{(s-1)}[\varphi ^{(0)}(\lambda
),\psi ^{(0)}(\mu )dx]  \label{3.34} \\
&=&\underset{S_{\pm }^{(s)}(\sigma _{(t;x)}^{(s-1)},\sigma
_{(t_{0};x_{0})}^{(s-1)})}{\int }Z^{(s)}[\varphi ^{(0)}(\lambda ),\psi
^{(0)}(\mu )dx],  \notag
\end{eqnarray}%
and, similarly, 
\begin{eqnarray}
&&\Omega _{(t;x)}^{\circledast }(\lambda ,\mu )-\Omega
_{(t_{0};x_{0})}^{\circledast }(\lambda ,\mu )=\int_{\partial
S_{(t;x)}^{(s)}}\bar{\Omega}^{(s-1),\intercal }[\varphi ^{(0)}(\lambda
),\psi ^{(0)}(\mu )dx]  \label{3.35} \\
&&-\int_{\partial S_{(t_{0};x_{0})}^{(s)}}\bar{\Omega}^{(s-1),\intercal
}[\varphi ^{(0)}(\lambda ),\psi ^{(0)}(\mu )dx]  \notag
\end{eqnarray}%
\begin{eqnarray*}
&=&\underset{S_{\pm }^{(s)}(\sigma _{(t;x)}^{(s-1)},\sigma
_{(t_{0};x_{0})}^{(s-1)})}{\int }d\bar{\Omega}^{(s-1),\intercal }[\varphi
^{(0)}(\lambda ),\psi ^{(0)}(\mu )dx] \\
&=&\underset{S_{\pm }^{(s)}(\sigma _{(t;x)}^{(s-1)},\sigma
_{(t_{0};x_{0})}^{(s-1)})}{\int }\bar{Z}^{(s-1),\intercal }[\varphi
^{(0)}(\lambda ),\psi ^{(0)}(\mu )dx],
\end{eqnarray*}%
where $\lambda ,\mu \in \Sigma ,$ and by definition, $s$-dimensional
surfaces $S_{+}^{(s)}(\sigma _{(t;x)}^{(s-1)},\sigma
_{(t_{0};x_{0})}^{(s-1)})$ and $S_{-}^{(s)}(\sigma _{(t;x)}^{(s-1)},\sigma
_{(t_{0};x_{0})}^{(s-1)})\subset C_{s-1}(M)$ are spanned smoothly without
self-intersection between two homological cycles $\sigma
_{(t;x)}^{(s-1)}=\partial S_{(t;x)}^{(s)}$ and \ $\sigma
_{(t_{0};x_{0})}^{(s-1)}:=\partial S_{(t_{0};x_{0})}^{(s)}\in C_{s-1}(M;%
\mathbb{C})$ in such a way that the boundary $\partial (S_{+}^{(s)}(\sigma
_{(t;x)}^{(s-1)},\sigma _{(t_{0};x_{0})}^{(s-1)})$ $\cup $ $%
S_{-}^{(s)}(\sigma _{(t;x)}^{(s-1)},\sigma
_{(t_{0};x_{0})}^{(s-1)}))=\oslash .$ Since the integral operator
expressions $\Omega _{(t_{0};x_{0})},\Omega _{(t_{0};x_{0})}^{\circledast
}:L_{2}^{(\rho )}(\Sigma ;\mathbb{C})\rightarrow L_{2}^{(\rho )}(\Sigma ;%
\mathbb{C})$ are at a fixed point $(t_{0};x_{0})\in M,$ \ evidently,
constant and assumed to be invertible, for extending the actions given (\ref%
{3.32}) on the whole Hilbert space $\mathcal{H\times H}^{\ast }$ one can
apply to them the classical constants variation approach, making use of the
expressions (\ref{3.35}). As a result, we obtain easily the following
Delsarte transmutation integral operator expressions%
\begin{equation}
\mathbf{\Omega }_{\pm }=\mathbf{1-}\int_{\Sigma \times \Sigma }d\rho (\xi
)d\rho (\eta )\tilde{\psi}(x;\xi )\Omega _{(t_{0};x_{0})}^{-1}(\xi ,\eta )%
\underset{S_{\pm }^{(s)}(\sigma _{(t;x)}^{(s-1)},\sigma
_{(t_{0};x_{0})}^{(s-1)})}{\int }Z^{(s)}[\varphi ^{(0)}(\eta ),\cdot ],
\label{3.36}
\end{equation}%
\begin{equation*}
\mathbf{\Omega }_{\pm }^{\circledast }=\mathbf{1-}\int_{\Sigma \times \Sigma
}d\rho (\xi )d\rho (\eta )\tilde{\varphi}(x;\eta )\Omega
_{(t_{0};x_{0})}^{\circledast ,-1}(\xi ,\eta )\underset{S_{\pm
}^{(s)}(\sigma _{(t;x)}^{(s-1)},\sigma _{(t_{0};x_{0})}^{(s-1)})}{\int }%
\text{ \ }\bar{Z}^{(s),\intercal }[\cdot ,\psi ^{(0)}(\xi )dx]
\end{equation*}%
for fixed pairs $(\varphi ^{(0)}(\xi ),\psi ^{(0)}(\eta ))\in \mathcal{H}%
_{0}^{\ast }\times \mathcal{H}_{0}$ and $(\tilde{\varphi}^{(0)}(\lambda ),%
\tilde{\psi}^{(0)}(\mu ))\in \mathcal{\tilde{H}}_{0}^{\ast }\times \mathcal{%
\tilde{H}}_{0},$ $\lambda ,\mu \in \Sigma ,$ being bounded invertible
integral operators of \ Volterra type on the whole space $\mathcal{H}\times 
\mathcal{H}^{\ast }.$ Applying the same arguments as in Section 1, one can
show also that respectively transformed sets of operators $\mathrm{\tilde{L}}%
_{j}:=\mathbf{\Omega }_{\pm }\mathrm{L}_{j}\mathbf{\Omega }_{\pm }^{-1},$ $j=%
\overline{1,r},$ and \ $\mathrm{\tilde{L}}_{k}^{\ast }:=\mathbf{\Omega }%
_{\pm }^{\circledast }\mathrm{L}_{k}^{\ast }\mathbf{\Omega }_{\pm
}^{\circledast ,-1},$ $k=\overline{1,r},$ prove to be purely differential
too. Thereby, one can formulate the following final theorem.

\begin{theorem}
The expressions (\ref{3.36}) are bounded invertible Delsarte transmutation
integral operators of Volterra type onto $\mathcal{H}\times \mathcal{H}%
^{\ast },$ transforming, respectively, given commuting sets of operators $%
\mathrm{L}_{j},$ $j=\overline{1,r},$ and their formally adjoint ones $%
\mathrm{L}_{k}^{\ast },$ $k$ $=\overline{1,r},$ into the pure differential
sets of operators $\mathrm{\tilde{L}}_{j}:=\mathbf{\Omega }_{\pm }\mathrm{L}%
_{j}\mathbf{\Omega }_{\pm }^{-1},$ $j=\overline{1,r},$ and \ $\mathrm{\tilde{%
L}}_{k}^{\ast }:=\mathbf{\Omega }_{\pm }^{\circledast }\mathrm{L}_{k}^{\ast }%
\mathbf{\Omega }_{\pm }^{\circledast ,-1},$ $k=\overline{1,r}.$ Moreover,
the suitably constructed closed subspaces $\mathcal{H}_{0}\subset \mathcal{H}
$ \ and $\mathcal{\tilde{H}}_{0}\subset \mathcal{H},$ such that the operator 
$\mathbf{\Omega }\in Aut(\mathcal{H})\cap \mathcal{B(H)}$ depend strongly on
the topological structure of the generalized cohomology groups $\mathcal{H}%
_{\Lambda (\mathcal{L}),-}^{0}(M)$ and $\mathcal{H}_{\Lambda (\mathcal{%
\tilde{L}}),-}^{0}(M),$ being parametrized by elements $S_{\pm
}^{(s)}(\sigma _{(t;x)}^{(s-1)},\sigma _{(t_{0};x_{0})}^{(s-1)})\in C_{s}(M;%
\mathbb{C}).$
\end{theorem}

5.7. Suppose now that all of differential operators $L_{j}:=L_{j}(x|\partial
),$ $j=\overline{1,r},$ considered above don't depend on the variable $t\in 
\mathrm{T}^{r}\subset \mathbb{R}_{+}^{r}.$ Then, evidently, one can take 
\begin{eqnarray*}
\mathcal{H}_{0} &:&=\{\psi _{\mu }^{(0)}(\xi )\in L_{2,-}(\mathbb{R}^{s};%
\mathbb{C}^{N}):L_{j}\psi _{\mu }^{(0)}(\xi )=\mu _{j}\psi _{\mu }^{(0)}(\xi
),\text{ \ }j=\overline{1,r}, \\
\text{\ }\psi _{\mu }^{(0)}(\xi )|_{\Gamma } &=&0,\text{ }\mu :=(\mu
_{1},...,\mu _{r})\in \sigma (\mathcal{\tilde{L}})\cap \bar{\sigma}(\mathcal{%
L}^{\ast }),\text{ }\xi \in \Sigma _{\sigma }\},
\end{eqnarray*}%
\begin{eqnarray*}
\mathcal{\tilde{H}}_{0} &:&=\{\tilde{\psi}_{\mu }^{(0)}(\xi )\in L_{2,-}(%
\mathbb{R}^{s};\mathbb{C}^{N}):\tilde{L}_{j}\tilde{\psi}_{\mu }^{(0)}(\xi
)=\mu _{j}\tilde{\psi}_{\mu }^{(0)}(\xi ),\text{ \ }j=\overline{1,r}, \\
\text{ \ }\tilde{\psi}_{\mu }^{(0)}(\xi )|_{\tilde{\Gamma}} &=&0,\text{ }\mu
:=(\mu _{1},...,\mu _{r})\in \sigma (\mathcal{\tilde{L}})\cap \bar{\sigma}(%
\mathcal{L}^{\ast }),\text{ }\xi \in \Sigma _{\sigma }\},
\end{eqnarray*}%
\begin{eqnarray}
\mathcal{H}_{0}^{\ast } &:&=\{\varphi _{\lambda }^{(0)}(\eta )\in L_{2,-}(%
\mathbb{R}^{s};\mathbb{C}^{N}):L_{j}^{\ast }\varphi _{\lambda }^{(0)}(\eta )=%
\bar{\lambda}_{j}\varphi _{\lambda }^{(0)}(\eta ),\text{ \ }j=\overline{1,r},
\label{3.37} \\
\text{\ \ }\varphi _{\lambda }^{(0)}(\eta )|_{\Gamma } &=&0,\text{ }\lambda
:=(\lambda _{1},...,\lambda _{r})\in \sigma (\mathcal{\tilde{L}})\cap \bar{%
\sigma}(\mathcal{L}^{\ast }),\text{ }\eta \in \Sigma _{\sigma }\},  \notag
\end{eqnarray}%
\begin{eqnarray*}
\mathcal{\tilde{H}}_{0}^{\ast } &:&=\{\tilde{\varphi}_{\lambda }^{(0)}(\eta
)\in L_{2,-}(\mathbb{R}^{s};\mathbb{C}^{N}):\tilde{L}_{j}^{\ast }\tilde{%
\varphi}_{\lambda }^{(0)}(\eta )=\bar{\lambda}_{j}\tilde{\varphi}_{\lambda
}^{(0)}(\eta ),\text{ \ }j=\overline{1,r}, \\
\text{\ \ }\tilde{\varphi}_{\lambda }^{(0)}(\eta )|_{\tilde{\Gamma}} &=&0,%
\text{ }\lambda :=(\lambda _{1},...,\lambda _{r})\in \sigma (\mathcal{\tilde{%
L}})\cap \bar{\sigma}(\mathcal{L}^{\ast }),\text{ }\eta \in \Sigma _{\sigma
}\}
\end{eqnarray*}%
and construct the corresponding Delsarte transmutation operators 
\begin{eqnarray}
\mathbf{\Omega }_{\pm } &=&\mathbf{1-}\underset{\sigma (\mathcal{\tilde{L}}%
)\cap \bar{\sigma}(\mathcal{L}^{\ast })}{\int }d\rho _{\sigma }(\lambda )%
\underset{\Sigma _{\sigma }\times \Sigma _{\sigma }}{\int }d\rho _{\Sigma
_{\sigma }}(\xi )d\rho _{\Sigma _{\sigma }}(\eta )  \label{3.38} \\
&&\times \underset{S_{\pm }^{(s)}(\sigma _{x}^{(s-1)},\sigma
_{x_{0}}^{(s-1)})}{\int dx}\tilde{\psi}_{\lambda }^{(0)}(\xi )\Omega
_{(x_{0})}^{-1}(\lambda ;\xi ,\eta )\bar{\varphi}_{\lambda }^{(0),\intercal
}(\eta )(\cdot )  \notag
\end{eqnarray}%
and 
\begin{eqnarray}
\mathbf{\Omega }_{\pm }^{\circledast } &=&\mathbf{1-}\underset{\sigma (%
\mathcal{\tilde{L}})\cap \bar{\sigma}(\mathcal{L}^{\ast })}{\int }d\rho
_{\sigma }(\lambda )\int_{\Sigma }d\rho _{\Sigma _{\sigma }}(\xi )d\rho
_{\Sigma _{\sigma }}(\eta )  \label{3.39} \\
&&\times \underset{S_{\pm }^{(s)}(\sigma _{x}^{(s-1)},\sigma
_{x_{0}}^{(s-1)})}{\int dx}\tilde{\varphi}_{\lambda }^{(0)}(\xi )\bar{\Omega}%
_{(x_{0})}^{\intercal ,-1}(\lambda ;\xi ,\eta )\times \bar{\psi}_{\lambda
}^{(0),\intercal }(\eta )(\cdot ),  \notag
\end{eqnarray}%
acting already in the Hilbert space $L_{2}(\mathbb{R}^{s};\mathbb{C}^{N}),$
where for any $(\lambda ;\xi ,\eta )\in (\sigma (\mathcal{\tilde{L}})\cap 
\bar{\sigma}(\mathcal{L}^{\ast }))\times \Sigma _{\sigma }^{2}$ kernels 
\begin{eqnarray}
\Omega _{(x_{0})}(\lambda ;\xi ,\eta ) &:&=\int_{\sigma
_{x_{0}}^{(s-1)}}\Omega ^{(s-1)}[\varphi _{\lambda }^{(0)}(\xi ),\psi
_{\lambda }^{(0)}(\eta )dx],  \label{3.39a} \\
\Omega _{(x_{0})}^{\circledast }(\lambda ;\xi ,\eta ) &:&=\int_{\sigma
_{x_{0}}^{(s-1)}}\bar{\Omega}^{(s-1),\intercal }[\varphi _{\lambda
}^{(0)}(\xi ),\psi _{\lambda }^{(0)}(\eta )dx]  \notag
\end{eqnarray}%
belong to $L_{2}^{(\rho )}(\Sigma _{\sigma };\mathbb{C})\times L_{2}^{(\rho
)}(\Sigma _{\sigma };\mathbb{C})$ for every $\lambda \in \sigma (\mathcal{%
\tilde{L}})\cap \bar{\sigma}(\mathcal{L}^{\ast })$ considered as a parameter$%
.$ Moreover, as $\partial \mathbf{\Omega }_{\pm }/\partial t_{j}=0,$ $j=%
\overline{1,r},$ one gets easily the set of differential expressions 
\begin{equation}
\mathcal{R}(\mathcal{\tilde{L}}):=\{\tilde{L}_{j}(x|\partial ):=\mathbf{%
\Omega }_{\pm }L_{j}(x|\partial )\mathbf{\Omega }_{\pm }^{-1}:j=\overline{1,r%
}\},  \label{3.40}
\end{equation}%
being a ring of commuting to each other differential operators acting in $\
L_{2}(\mathbb{R}^{s};\mathbb{C}^{N}),$ generated by the corresponding
initial ring $\mathcal{R}(\mathcal{L}).$

Thus we have described above a ring $\mathcal{R}(\mathcal{\tilde{L}})$ of
commuting to each other multi-dimensional differential operators, generated
by an initial ring $\mathcal{R}(\mathcal{L}).$ This problem in the
one-dimensional case was before treated in detail and effectively solved in 
\cite{No,Kr} \ by means of algebro-geometric\ and inverse spectral transform
techniques. Our approach gives another look at this problem in
multidimension and is of special interest due to its clear and readable
dependence on dimension of differential operators.

\section{A special case: soliton theory aspect}

6.1. Consider our de Rham-Hodge theory of a commuting set $\mathcal{L}$ of
two differential operators in a Hilbert space $\mathcal{H}:=L_{2}(\mathrm{T}%
^{2};H),$ $H:=L_{2}(\mathbb{R}^{s};\mathbb{C}^{N}),$ for the special case
when $M:=\mathrm{T}^{2}\times \mathbb{\bar{R}}^{s}$ and 
\begin{equation}
\mathcal{L}:=\{\mathrm{L}_{j}:=\partial /\partial t_{j}-L_{j}(t;x|\partial
):t_{j}\in \text{\ }\mathrm{T}_{j}:=[0,T_{j})\subset \mathbb{R}_{+},\text{ }%
j=\overline{1,2}\},  \notag
\end{equation}%
where, by definition, $\mathrm{T}^{2}:=\mathrm{T}_{1}\times \mathrm{T}_{2},$ 
\begin{equation}
L_{j}(t;x|\partial ):=\sum_{|\alpha |=0}^{n(L_{j})}a_{\alpha
}^{(j)}(t;x)\partial ^{|\alpha |}/\partial x^{\alpha }  \label{4.2}
\end{equation}%
with coefficients $a_{\alpha }^{(j)}\in C^{1}(\mathrm{T}^{2};S(\mathbb{R}%
^{s};End\mathbb{C}^{N})),$ $\alpha \in \mathbb{Z}_{+}^{s},$ $|\alpha |=%
\overline{0,n(L_{j})},$ $j=\overline{1,2}.$ The corresponding scalar product
is given now as 
\begin{equation}
(\varphi ,\psi ):=\int_{\mathrm{T}^{2}}dt\int_{\mathbb{R}^{s}}dx<\varphi
,\psi >  \label{4.3}
\end{equation}%
for any pair $(\varphi ,\psi )\in \mathcal{H}\times \mathcal{H}$ and the
generalized external differential 
\begin{equation}
d_{\mathcal{L}}:=\sum_{j=1}^{2}dt_{j}\wedge \mathrm{L}_{j},  \label{4.4}
\end{equation}%
where one assumes that for all $t\in \mathrm{T}^{2}$ and $x\in \mathbb{R}%
^{s} $ the commutator 
\begin{equation}
\lbrack \mathrm{L}_{1},\mathrm{L}_{2}]=0.  \label{4.5}
\end{equation}%
Tis means, obviously, that the corresponding de Rham-Hodge-Skrypnik co-chain
complexes%
\begin{eqnarray}
\mathcal{H} &\rightarrow &\Lambda ^{0}(M;\mathcal{H)}\overset{d_{\mathcal{L}}%
}{\rightarrow }\Lambda ^{1}(M;\mathcal{H)}\overset{d_{\mathcal{L}}}{%
\rightarrow }...\overset{d_{\mathcal{L}}}{\rightarrow }\Lambda ^{m}(M;%
\mathcal{H)}\overset{d_{\mathcal{L}}}{\rightarrow }0,  \label{4.6} \\
\mathcal{H} &\rightarrow &\Lambda ^{0}(M;\mathcal{H)}\overset{d_{\mathcal{L}%
}^{\ast }}{\rightarrow }\Lambda ^{1}(M;\mathcal{H)}\overset{d_{\mathcal{L}%
}^{\ast }}{\rightarrow }...\overset{d_{\mathcal{L}}^{\ast }}{\rightarrow }%
\Lambda ^{m}(M;\mathcal{H)}\overset{d_{\mathcal{L}}^{\ast }}{\rightarrow }0 
\notag
\end{eqnarray}%
are exact. Define now due to (\ref{3.10}) and (\ref{3.31}) the closed
subspaces $H_{0}^{\circledast }$ $\ \ $and $H_{0}\subset H_{-}$ as follows:%
\begin{eqnarray}
\mathcal{H}_{0} &:&=\{\psi ^{(0)}(\lambda ;\eta )\in \mathcal{H}_{\Lambda (%
\mathcal{L}),-}^{0}(M):  \label{4.7} \\
\partial \psi ^{(0)}(\lambda ;\eta )/\partial t_{j} &=&L_{j}(t;x|\partial
)\psi ^{(0)}(\lambda ;\eta ),\text{ }j=\overline{1,2},  \notag \\
\psi ^{(0)}(\lambda ;\eta )|_{t=t_{0}} &=&\psi _{\lambda }(\eta )\in H_{-},%
\text{ }\psi ^{(0)}(\lambda ;\eta )|_{\Gamma }=0,  \notag \\
(\lambda ;\eta ) &\in &\Sigma \subset (\sigma (\mathcal{L})\cap \bar{\sigma}(%
\mathcal{L}^{\ast }))\times \Sigma _{\sigma }\},  \notag
\end{eqnarray}%
\begin{eqnarray*}
\mathcal{H}_{0}^{\ast } &:&=\{\mathbb{\varphi }^{(0)}(\lambda ;\eta )\in 
\mathcal{H}_{\Lambda (\mathcal{L}),-}^{0}(M): \\
-\partial \mathbb{\varphi }^{(0)}(\lambda ;\eta )/\partial t_{j}
&=&L_{j}^{\ast }(t;x|\partial )\mathbb{\varphi }^{(0)}(\lambda ;\eta ),\text{
}j=\overline{1,2}, \\
\mathbb{\varphi }^{(0)}(\lambda ;\eta )|_{t=t_{0}} &=&\mathbb{\varphi }%
_{\lambda }(\eta )\in H_{-},\text{ }\mathbb{\varphi }^{(0)}(\lambda ;\eta
)|_{\Gamma }=0, \\
\text{ }(\lambda ;\eta ) &\in &\Sigma \subset (\sigma (\mathcal{L})\cap \bar{%
\sigma}(\mathcal{L}^{\ast }))\times \Sigma _{\sigma }\}
\end{eqnarray*}%
for some hypersurface $\Gamma \subset M$ and a "spectral" degeneration set \ 
$\Sigma _{\sigma }\in \mathbb{C}^{p-1}.$ By means of subspaces (\ref{4.7})
one can now proceed to construction of Delsarte transmutation operators $%
\mathbf{\Omega }_{\pm }:H\leftrightarrows H$ in the general form like (\ref%
{3.39}) with kernels $\Omega _{(t_{0};x_{0})}(\lambda ;\xi ,\eta )\in
L_{2}^{(\rho )}(\Sigma _{\sigma };\mathbb{C})\otimes L_{2}^{(\rho )}(\Sigma
_{\sigma };\mathbb{C})$ for every $\lambda \in \sigma (\mathcal{L})\cap \bar{%
\sigma}(\mathcal{L}^{\ast }),$ being defined as 
\begin{eqnarray}
\Omega _{(t_{0};x_{0})}(\lambda ;\xi ,\eta ) &:&=\int_{\sigma
_{(t_{0};x_{0})}^{(s-1)}}\Omega ^{(s-1)}[\mathbb{\varphi }^{(0)}(\lambda
;\xi ),\psi ^{(0)}(\lambda ;\eta )dx],  \label{4.8} \\
\Omega _{(t_{0};x_{0})}^{\circledast }(\lambda ;\xi ,\eta ) &:&=\int_{\sigma
_{(t_{0};x_{0})}^{(s-1)}}\bar{\Omega}^{(s-1),\intercal }[\mathbb{\varphi }%
^{(0)}(\lambda ;\xi ),\psi ^{(0)}(\lambda ;\eta )dx]  \notag
\end{eqnarray}%
for all $(\lambda ;\xi ,\eta )\in (\sigma (\mathcal{L})\cap \bar{\sigma}(%
\mathcal{L}^{\ast }))\times \Sigma _{\sigma }^{2}.$ As a result one gets for
the corresponding product $\rho :=\rho _{\sigma }\odot \rho _{\Sigma
_{\sigma }^{2}}$ such integral expressions:%
\begin{eqnarray*}
\mathbf{\Omega }_{\pm } &=&\mathbf{1-}\underset{\sigma (\mathcal{L})\cap 
\bar{\sigma}(\mathcal{L}^{\ast })}{\int }d\rho _{\sigma }(\lambda )\underset{%
\Sigma _{\sigma }\times \Sigma _{\sigma }}{\int }d\rho _{\Sigma _{\sigma
}}(\xi )d\rho _{\Sigma _{\sigma }}(\eta ) \\
&&\times \underset{S_{\pm }^{(s)}(\sigma _{(t_{0};x)}^{(s-1)},\sigma
_{(t_{0};x_{0})}^{(s-1)})}{\int dx}\tilde{\psi}^{(0)}(\lambda ;\xi )\Omega
_{(t_{0};x_{0})}^{-1}(\lambda ;\xi ,\eta )\bar{\varphi}^{(0),\intercal
}(\lambda ;\eta )(\cdot ),
\end{eqnarray*}%
\begin{eqnarray}
\mathbf{\Omega }_{\pm }^{\circledast } &=&\mathbf{1-}\underset{\sigma (%
\mathcal{L})\cap \bar{\sigma}(\mathcal{L}^{\ast })}{\int }d\rho _{\sigma
}(\lambda )\underset{\Sigma _{\sigma }\times \Sigma _{\sigma }}{\int }d\rho
_{\Sigma _{\sigma }}(\xi )d\rho _{\Sigma _{\sigma }}(\eta )  \label{4.9} \\
&&\times \underset{S_{\pm }^{(s)}(\sigma _{(t_{0};x)}^{(s-1)},\sigma
_{(t_{0};x_{0})}^{(s-1)})}{\int dx}\tilde{\varphi}_{\lambda }^{(0)}(\xi )%
\bar{\Omega}_{(t_{0};x_{0})}^{\intercal ,-1}(\lambda ;\xi ,\eta )\times \bar{%
\psi}^{(0),\intercal }(\lambda ;\eta )(\cdot ),  \notag
\end{eqnarray}%
where $S_{+}^{(s)}(\sigma _{(t_{0};x)}^{(s-1)},\sigma
_{(t_{0};x_{0})}^{(s-1)})\in C_{s}(M;\mathbb{C})$ is some smooth $s$%
-dimensional surface spanned between two homological cycles $\sigma
_{(t_{0};x)}^{(s-1)}$ \ and $\sigma _{(t_{0};x_{0})}^{(s-1)}\in \mathcal{K}%
(M)$ and $S_{-}^{(s)}(\sigma _{(t_{0};x)}^{(s-1)},\sigma
_{(t_{0};x_{0})}^{(s-1)})\in C_{s}(M;\mathbb{C})$ is its smooth counterpart
such that $\partial (S_{+}^{(s)}(\sigma _{(t_{0};x)}^{(s-1)},\sigma
_{(t_{0};x_{0})}^{(s-1)})\cup S_{-}^{(s)}(\sigma _{(t_{0};x)}^{(s-1)},\sigma
_{(t_{0};x_{0})}^{(s-1)}))=\oslash .$ Concerning the related results of
Section 3 one can construct from (\ref{4.9}) the corresponding factorized
Fredholm operators $\mathbf{\Omega }$ and $\ \mathbf{\Omega }^{\circledast
}:H\rightarrow H,$ $H=L_{2}(\mathbb{R};\mathbb{C}^{N}),$ as follows:%
\begin{equation}
\mathbf{\Omega :=\Omega }_{+}^{-1}\mathbf{\Omega }_{-},\text{ \ \ \ }\mathbf{%
\Omega }^{\circledast }\mathbf{:=\Omega }_{+}^{\circledast -1}\mathbf{\Omega 
}_{-}^{\circledast }.  \label{4.10}
\end{equation}%
It is also important to notice here that kernels $\hat{K}_{\pm }(\mathbf{%
\Omega })$ and $\hat{K}_{\pm }(\mathbf{\Omega }^{\circledast })\in
H_{-}\otimes H_{-}$ satisfy exactly the generalized \cite{Be} determining
equations\ in the following tensor form 
\begin{eqnarray}
(\mathcal{\tilde{L}}_{ext}\otimes \mathbf{1)}\hat{K}_{\pm }(\mathbf{\Omega }%
) &=&(\mathbf{1\otimes }\mathcal{L}_{ext}^{\ast }\mathbf{)}\hat{K}_{\pm }(%
\mathbf{\Omega }),  \label{4.11} \\
(\mathcal{\tilde{L}}_{ext}^{\ast }\otimes \mathbf{1)}\hat{K}_{\pm }(\mathbf{%
\Omega }^{\circledast }) &=&(\mathbf{1\otimes }\mathcal{L}_{ext}\mathbf{)}%
\hat{K}_{\pm }(\mathbf{\Omega }^{\circledast }).  \notag
\end{eqnarray}%
Since, evidently, $supp\hat{K}_{+}(\mathbf{\Omega })\cap supp\hat{K}_{-}(%
\mathbf{\Omega })=\oslash $ and $\ supp\hat{K}_{+}(\mathbf{\Omega }%
^{\circledast })\cap supp\hat{K}_{-}(\mathbf{\Omega }^{\circledast
})=\oslash ,$ one derives from results \cite{My,GPSP,PSP} \ that
corresponding Gelfand-Levitan-Marchenko equations 
\begin{eqnarray}
\hat{K}_{+}(\mathbf{\Omega })+\hat{\Phi}(\mathbf{\Omega )+}\hat{K}_{+}(%
\mathbf{\Omega })\ast \hat{\Phi}(\mathbf{\Omega )} &\mathbf{=}&\hat{K}_{-}(%
\mathbf{\Omega }),  \label{4.12} \\
\hat{K}_{+}(\mathbf{\Omega }^{\circledast })+\hat{\Phi}(\mathbf{\Omega }%
^{\circledast }\mathbf{)+}\hat{K}_{+}(\mathbf{\Omega }^{\circledast })\ast 
\hat{\Phi}(\mathbf{\Omega }^{\circledast }\mathbf{)} &\mathbf{=}&\hat{K}_{-}(%
\mathbf{\Omega }^{\circledast }),  \notag
\end{eqnarray}%
where, by definition, $\mathbf{\Omega :}=\mathbf{1}+\hat{\Phi}(\mathbf{%
\Omega }),$ $\mathbf{\Omega }^{\circledast }:=1+\hat{\Phi}(\mathbf{\Omega }%
^{\circledast }),$ can be solved \cite{My,GK} in the space $\mathcal{B}%
_{\infty }^{\pm }(H)$ for kernels $\hat{K}_{\pm }(\mathbf{\Omega })$ and $%
\hat{K}_{\pm }(\mathbf{\Omega }^{\circledast })\in H_{-}\otimes H_{-}$
depending parametrically on $t\in \mathrm{T}^{2}.$ Thereby, Delsarte
transformed differential operators $\mathrm{\tilde{L}}_{j}:\mathcal{H}%
\rightarrow \mathcal{H},$ $j=\overline{1,2},$ will be, evidently, commuting
to each other too, satisfying the following operator relationships: 
\begin{equation}
\mathrm{\tilde{L}}_{j}=\partial /\partial t_{j}-\mathbf{\Omega }_{\pm }L_{j}%
\mathbf{\Omega }_{\pm }^{-1}-(\partial \mathbf{\Omega }_{\pm }/\partial
t_{j})\mathbf{\Omega }_{\pm }^{-1}:=\partial /\partial t_{j}-\tilde{L}_{j},%
\text{ }  \label{4.13}
\end{equation}%
where operator expressions for $\tilde{L}_{j}\in \mathcal{L}(H),$ $j=%
\overline{1,2},$ prove to be purely differential. The latter property makes
it possible to construct some nonlinear in general partial differential
equations on coefficients of differential operators (\ref{4.13}) and solve
them by means of the standard procedures either of inverse spectral \cite%
{No,Ma,LS,Le} or the Darboux-Backlund \cite{MS,SPS,PSPS} transforms,
producing a wide class of exact soliton like solutions. Another not simple
and very interesting aspect of the approach devised in this paper concerns
regular algorithms of treating differential operator expressions depending
on a "spectral" parameter $\lambda \in \mathbb{C},$\ \ which was just a
little recently discussed in \cite{GPSP,PSP}.

\section{Conclusion}

The review done above presents recent results \ devoted to the development
of \ the De Rham-Hodge-Skrypnik theory \cite{DeR,DeR1,Te,Sk,Wa} of special
differential complexes which give rise to effective analytical expressions
for the corresponding Delsarte transmutation Volterra type operators in a
given Hilbert space $\mathcal{H}.$ In particular, it was shown that they can
be effectively applied to studying the integral operator structure of
Delsarte transmutation operators for polynomial pencils of differential
operators in $\mathcal{H}$\ \ having many applications both in spectral
theory of such \ multidimensional operator pencils and in soliton theory 
\cite{Ni,FT,No,GPSP,PM} of multidimensional integrable dynamical systems on
functional manifolds, being very important for diverse applications in
modern mathematical physics. If one considers a differential operator $L:%
\mathcal{H}\rightarrow \mathcal{H}$ and assumes that its spectrum $\sigma
(L) $ consists of the discrete $\sigma _{d}(L)$ and continuous $\sigma
_{c}(L)$ parts, by means of the general form of the Delsarte transmutation
operators obtained in Sections 4 and 5 one can construct a new more
complicated differential operator $\tilde{L}:=\mathbf{\Omega }_{\pm }L%
\mathbf{\Omega }_{\pm }^{-1}$ in $\mathcal{H},$ such that its continuous
spectrum $\sigma _{c}(\tilde{L})=\sigma _{c}(L)$ but $\sigma _{d}(L)\neq
\sigma _{d}(\tilde{L}).$ Thereby these Delsarte transformed operators can be
effectively used for both studying generalized spectral properties of
differential operators and operator pencils \cite{Be,Fa,Ma,LS,BS,DS,Su} and
constructing a wide class of nontrivial differential operators with a
prescribed spectrum as it was done \-\cite{Ma,No,DSa} in one dimension.

As it was shown before in \cite{Fa,Ni} for the two-dimensional Dirac and
three-dimensional perturbed Laplace operators, the kernels of the
corresponding Delsarte transmutation operator satisfy some special of
Fredholm type linear integral equations called the Gelfand-Levitan-Marchenko
ones, which are of very importance for solving the corresponding inverse
spectral problem and having many applications in modern mathematical
physics. Such equations can be easily constructed for our multidimensional
case too, thereby making it possible to pose the corresponding inverse
spectral problem for describing a wide class of multidimensional operators
with a priori given spectral characteristics. Also, similar to \cite%
{Ni,PM,No,Ma}, one can use such results for studying so called completely
integrable nonlinear evolution equations, especially for constructing by
means of special Darboux type transformations \cite{MS,SPS,PSPS} their exact
solutions like solitons and many others. Such an activity is now in progress
and the corresponding results will be published later.

\section{Acknowledgements.}

One of authors (A.P.) cordially thanks Prof. I.V. Skrypnik (IM, Kyiv and
IAM, Donetsk) for fruitful discussions of \ some aspects of the De Rham
-Hodge-Skrypnik theory and its applications presented in the article. The
authors are appreciated also very much to Profs. L.P. Nizhnik (IM of NAS,
Kyiv),\ Holod P.I. (UKMA, Kyiv), T. Winiarska (IM, Politechnical University,
Krakow), A. Pelczar and J. Ombach (Jagiellonian University, Krakow), J.
Janas (Institute of Mathematics of PAN, Krakow), Z. Peradzynski (Warsaw
University, Warsaw) and D.L. Blackmore (NJ Institute of \ Technology,
Newark, NJ, USA) for valuable comments on diverse problems related with
results presented in the article. The last but not least thanks is addressed
to our friends Profs. V.V. Gafiychuk (IAPMM, Lviv) and Ya. V. Mykytiuk
(I.Ya. Franko National University, Lviv) for the permanent support and help
in editing the article.

\bigskip

\end{document}